# Time-spectral control of accidental coincidences in daylight entanglement-based free-space QKD

Jiyoung Moon, Yonggi Jo, Zaeill Kim, Yong Sup Ihn and Nam Hun Park*
Agency for Defense Development, Yuseong P.O. Box 35, Daejeon 34186, Republic of Korea
Corresponding author: Nam Hun Park
Email: pnh@add.re.kr

## Abstract

Daylight entanglement-based free-space quantum key distribution (QKD) is limited by accidental coincidences from receiver-admitted background light. We develop and experimentally validate a receiver-level framework anchored to measured singles rates and reference coincidence components, linking receiver bandwidth, accepted temporal width, and background-noise density to Bob singles, sifted-key rate, error rate, and quantum bit error rate (QBER) in telecom-wavelength BBM92 QKD. Indoor sweeps show that the sifted-key rate saturates near the source-matched bandwidth, whereas broader bandwidth or higher background mainly increases accidental contamination. Increasing the accepted temporal width leaves Bob singles nearly unchanged but directly raises QBER by enlarging the random-overlap probability. A two-dimensional design map calculated from the model shows that the temporal-window margin contracts rapidly with increasing background-to-signal ratio, while the bandwidth margin remains comparatively broad near source-matched filtering. A rooftop experiment demonstrates daylight operation over a 10 m free-space link, yielding a mean sifted-key rate of 2,811 cps and a mean QBER of 4.43%. This framework provides a quantitative basis for choosing receiver bandwidth and temporal acceptance to meet a target QBER under specified background conditions.

## 1. Introduction

Quantum key distribution (QKD) enables the distribution of cryptographic keys with information-theoretic security and has become a central platform in quantum communications [1-4]. Among the major QKD protocols, entanglement-based schemes such as E91 and BBM92 are especially attractive for free-space implementation because their security is directly linked to measured quantum correlations, while coincidence measurements allow useful signal events to be distinguished from uncorrelated background counts under lossy and noisy propagation. These schemes are also naturally compatible with passive-basis receivers, networked architectures, and satellite-oriented links [1,2,5-9]. Free-space QKD is therefore of particular interest in settings where fiber deployment is impractical or where mobile, temporary, long-distance, or space-ground quantum links are required [5-9]. A central challenge for daylight free-space QKD is background light. Solar radiation and atmospheric scattering increase receiver singles rates, thereby raising accidental coincidences between otherwise independent detection events. Daylight operation therefore depends critically on receiver-side filtering in both the spectral and temporal domains: spectral filtering suppresses out-of-band background photons, whereas temporal filtering limits accepted temporal overlap of uncorrelated detections. Because the usable signal bandwidth is constrained by the source spectrum, spectral filtering must be matched to the source bandwidth rather than narrowed indefinitely, while the accepted temporal width remains a direct lever for suppressing random coincidences. Previous studies have demonstrated long-distance free-space and entanglement-based QKD links [10-12], daylight and satellite-oriented operation [13-16], daylight entanglement-based and metropolitan free-space implementations [17,18], daylight operation and filtering studies [19-21], and system-level design approaches for free-space QKD [22].

Previous studies have developed quantitative models for optimizing free-space and entanglement-based QKD under loss and background noise [22–25]. For a pulsed telecom-band BBM92 receiver, a practical design question is how to determine the combinations of receiver bandwidth and accepted temporal width that meet a target QBER at a given background-to-signal ratio. Spectral filtering determines both the admitted signal spectrum and the background spectral density, whereas temporal filtering directly controls the probability that uncorrelated detection events are accepted as coincidences. Once the receiver bandwidth is approximately matched to the source spectrum, additional spectral bandwidth yields limited useful signal gain but continues to admit background photons. This makes the available temporal-window margin a critical receiver-design quantity under increasing background-to-signal ratio.

Here we address this question using a receiver-level rate model anchored to measured singles rates and reference true- and accidental-coincidence components. Using a polarization-entangled photon-pair source, we test the same model against Bob total singles, sifted-key rate, error rate, and QBER in separate indoor sweeps of receiver bandwidth, accepted temporal width, and background-noise density. We then use the model validated against these sweeps to calculate a two-dimensional design map identifying bandwidth–window combinations compatible with selected QBER levels under different background-to-signal conditions. A rooftop experiment over a 10 m free-space link provides a complementary demonstration of daylight operation. The contribution is an experimentally supported link from measured receiver counts to the accidental budget and calculated operating margins, providing a quantitative basis for selecting spectral and temporal filtering settings in daylight entanglement-based BBM92 QKD.

## 2. Results

### 2.1 Controlled indoor testbed and source/background characterization

Figure 1 shows the controlled indoor testbed used to investigate receiver-side time-spectral filtering in entanglement-based free-space BBM92 QKD. Unlike the rooftop experiment, the indoor validation setup did not include a free-space quantum channel; instead, the source output was delivered directly through fiber to the receiver-side testbed. The idler arm routed to Bob entered an indoor emulated quantum-channel section, indicated in Fig. 1 by a dashed-dotted box, in which amplified spontaneous emission (ASE) noise was passed through a fixed attenuator and a variable optical attenuator (VOA) and then combined with the Bob-side quantum channel through a 99:1 fiber coupler before receiver-side spectral filtering. Alice and Bob used identical polarization-analysis modules (see Methods), and all detection events were recorded with a superconducting nanowire single-photon detector (SNSPD)- and time-tagger-based BBM92 analysis system. A coincidence histogram acquired with custom software was used to set the relative channel delays in both the indoor and outdoor experiments.

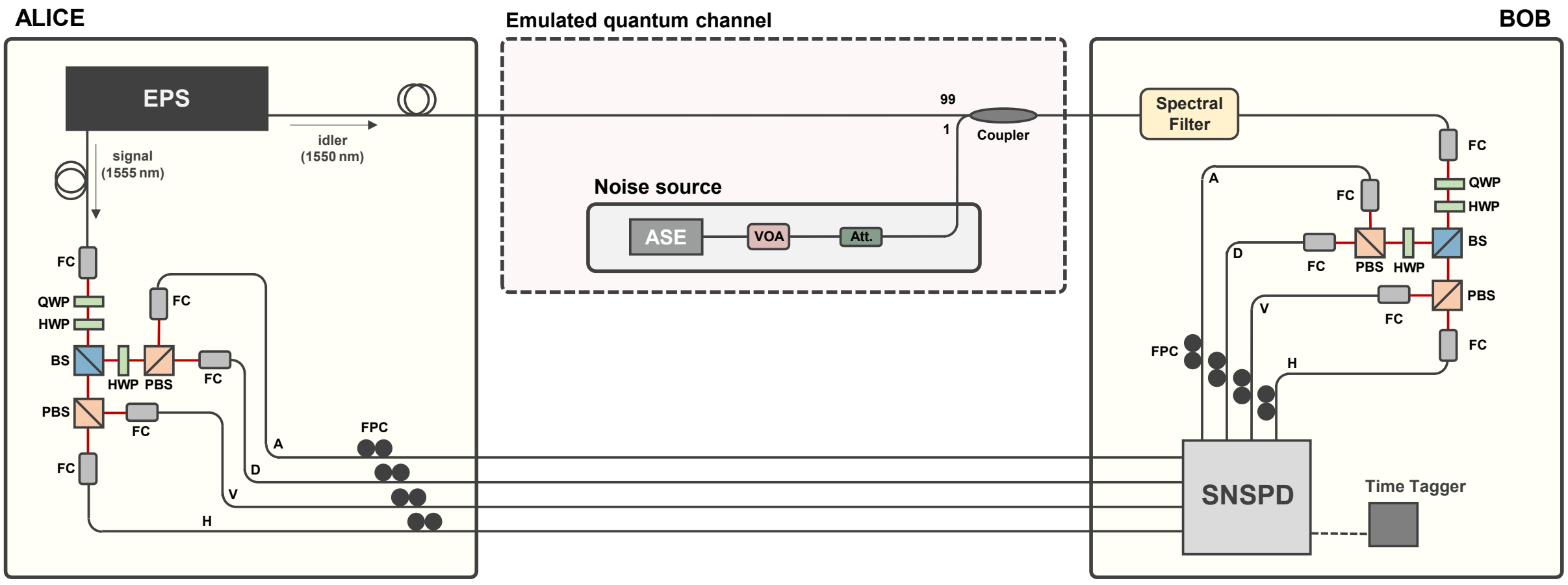

Figure 1. Controlled indoor testbed used for the time-spectral filtering study. A polarization-entangled photon-pair source generates signal and idler photons at 1555 nm and 1550 nm, respectively. In the indoor validation setup, the source output is delivered directly through fiber to the receiver-side testbed rather than through a free-space quantum channel. The signal photon is sent to Alice, whereas the idler photon enters the indoor emulated quantum-channel section, shown as a dashed-dotted box, where ASE background light passes through a fixed attenuator (ATT) and a VOA before being combined with the Bob-side channel through a 99:1 fiber coupler, spectrally filtered, and delivered to Bob. This indoor emulated quantum-channel section is the part replaced by the actual rooftop transmitter and receiver modules in the outdoor experiment. Alice and Bob use identical polarization-analysis modules in which a QWP, an HWP, and a BS separate the H/V and D/A outputs; the D/A branch contains an HWP set at 22.5° before a PBS, and all PBS outputs are fiber-coupled. The resulting eight channels are detected by an SNSPD system and analyzed using a time tagger with custom BBM92 software. Abbreviations: EPS, entangled photon source; ASE, amplified spontaneous emission; VOA, variable optical attenuator; ATT, fixed optical attenuator; QWP, quarter-wave plate; HWP, half-wave plate; BS, beam splitter; PBS, polarizing beam splitter; FC, fiber collimator; FPC, fiber polarization controller; SNSPD, superconducting nanowire single-photon detector.

The key temporal and spectral scales of the system are summarized in Fig. 2. Figure 2(a) shows a representative detector-plane coincidence histogram measured under the low-background reference condition over the delay range from -15 ns to 15 ns, with peaks separated by 10 ns, corresponding to the 100 MHz repetition rate. Figures 2(b) and 2(c) expand the main coincidence peak and the neighboring side peak, respectively; the main-peak full width at half maximum (FWHM) is 52 ps. From the detector-plane 16-channel summed coincidence histogram measured under the low-background reference condition, we obtained the reference true-coincidence component 19,893 cps and source-originated accidental-coincidence component 834 cps. Figure 2(d) shows the measured source spectrum, centered at 1550.07 nm with a FWHM of 0.64 nm, together with the super-Gaussian fit; the measured spectrum is well described by the super-Gaussian fit. Figure 2(e) shows the measured ASE background spectrum together with the transmission of the indoor filter, with ASE intensity plotted on the left y-axis (green) and filter transmission on the right y-axis (black). The indoor receiver filter used for this reference characterization was treated as having a constant insertion loss of 3.0 dB over the indoor bandwidth sweep. Together, these experimentally measured reference quantities defined the low-background input condition of the quantitative framework. Additional measured spectral characterization of the background and receiver filters is provided in Supplementary Note 1.

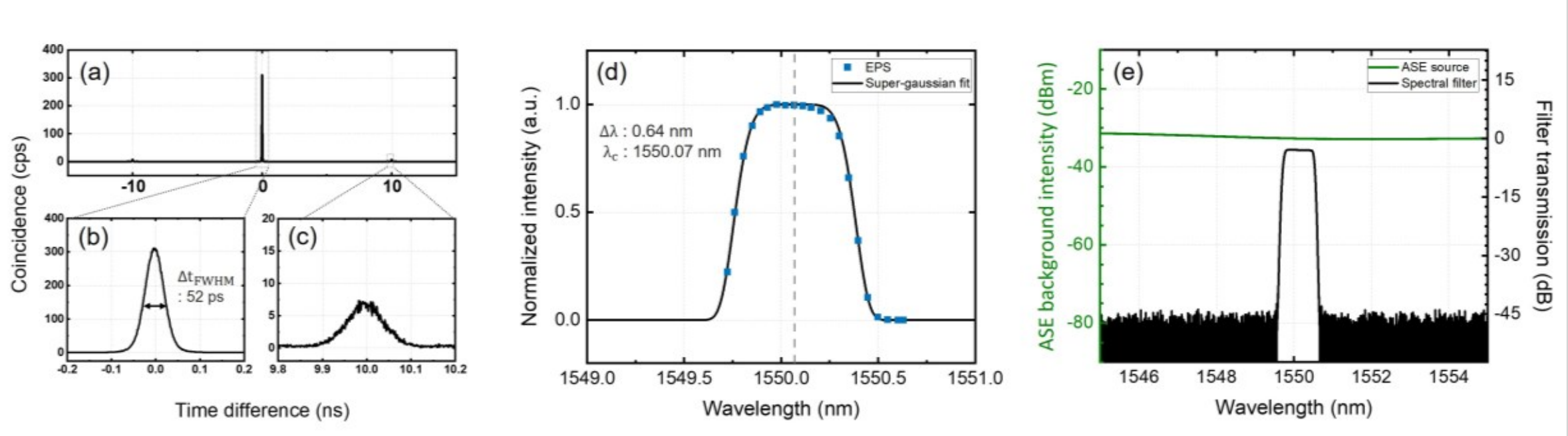

Figure 2. Temporal and spectral characterization used in the quantitative framework. (a) Representative coincidence histogram over the delay range from −15 ns to 15 ns, showing peaks separated by 10 ns for the 100 MHz repetition rate. (b) Expanded view of the main coincidence peak, with a FWHM of 52 ps. (c) Expanded view of the neighboring side peak at 10 ns. The detector-plane 16-channel summed coincidence histogram was used to obtain the reference true-coincidence component 19,893 cps and source-originated accidental-coincidence component 834 cps under the low-background reference condition. (d) Measured source spectrum, centered at 1550.07 nm with a FWHM of 0.64 nm, together with the super-Gaussian fit. (e) Measured ASE background spectrum and transmission profile of the indoor filter used for the reference characterization; the left y-axis (green) indicates ASE intensity in dBm and the right y-axis (black) indicates filter transmission, with spectral bandwidth 0.8 nm and insertion loss 3.0 dB.

## 2.2 Quantitative model for time-spectral filtering under background noise

To describe the measured QKD performance under background illumination, we distinguish coincidence components before basis sifting, $C_{\mathrm{i}}$, basis-sifted rates before multi-click rejection, $R_{\mathrm{i}}^{\mathrm{pre}}$, and post-rejection effective sifted rates, $R_{\mathrm{i}}$. The final sifted-key rate is written as $R_{\mathrm{sift}} = R_{\mathrm{true}} + R_{\mathrm{acc}}$, where $R_{\mathrm{acc}} = R_{\mathrm{acc,src}} + R_{\mathrm{acc,bg}}$. The pre-sifting background-induced accidental component is evaluated as $C_{\mathrm{acc,bg}} = S_{\mathrm{A,tot}}^{\mathrm{obs}} S_{\mathrm{B,bg}}^{\mathrm{obs}} T_{\mathrm{win}}$. Basis sifting and multi-click rejection then yield $R_{\mathrm{acc,bg}} = P_{\mathrm{keep}} p_{\mathrm{match}} C_{\mathrm{acc,bg}}$. The true-coincidence component is modeled using an effective spectral-truncation factor determined from the source and receiver bandwidths together with the temporal acceptance. Full definitions are provided in Supplementary Note 2. The model uses detector-plane reference quantities obtained from the measured 16-channel summed coincidence histogram and detector-plane singles rates under the low-background reference condition. These measured detector-plane anchors were internally converted to a common source-output-equivalent reference only to propagate variable receiver-side losses and filtering, and were then mapped back to detector-plane rates for comparison with the experiment, as detailed in Supplementary Note 2. From the resulting admitted Bob singles and coincidence budget, we obtain the model predictions compared with the measured Bob total singles, sifted-key rate, error rate, and QBER in Fig. 3.

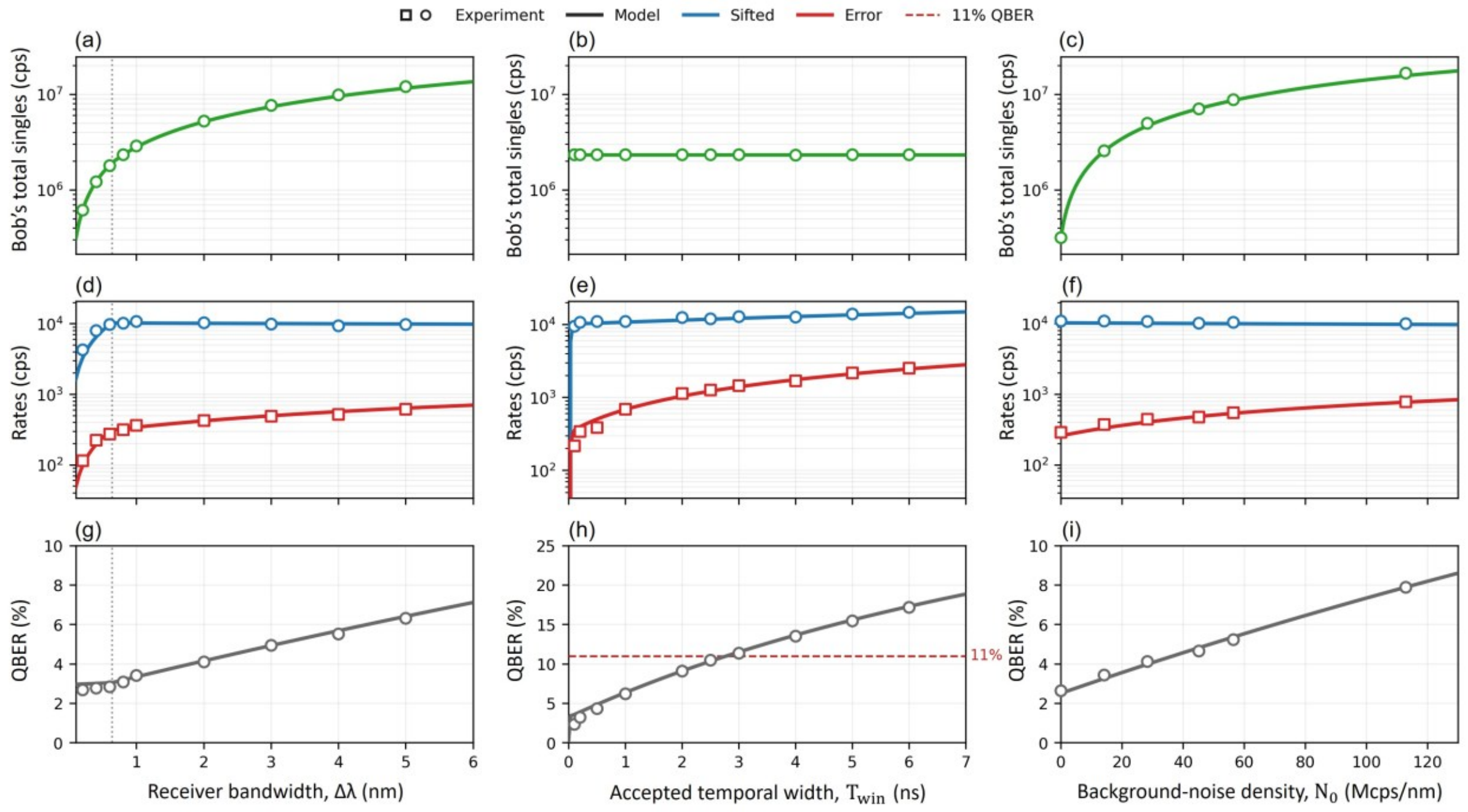

Figure 3. Controlled indoor validation of receiver-side time-spectral filtering effects. (a-c) Measured and calculated Bob total singles, defined as the sum of the H, V, D, and A count rates at Bob, as functions of receiver bandwidth $\Delta\lambda$, accepted temporal width $T_{\mathrm{win}}$, and background-noise density $N_0$, respectively. (d-f) Measured and calculated sifted-key and error rates. (g-i) Corresponding measured and calculated QBER. Fixed conditions are $N_0$ = 12.8 Mcps/nm and $T_{\mathrm{win}}$ = 0.2 ns for (a,d,g), $N_0$ = 12.8 Mcps/nm and $\Delta\lambda$=0.8 nm for (b,e,h), and $T_{\mathrm{win}}$ = 0.2 ns and $\Delta\lambda$ = 0.8 nm for (c,f,i). Markers denote experimental data and solid lines denote model results. Color coding is green for Bob total singles, blue and red for sifted-key and error rates, respectively, and dark gray for QBER; dashed red lines mark the 11% QBER benchmark. Model curves evaluate the background-induced accidental contribution using observed detector-plane singles after dead-time saturation and include the counted-and-discarded multi-click rejection correction described in Supplementary Note 2.

## 2.3 Experimental validation across bandwidth, window, and noise sweeps

Figure 3 compares measured Bob total singles, sifted-key rate, error rate, and QBER with the corresponding model predictions for systematic sweeps of receiver bandwidth, accepted temporal width, and background-noise density, thereby testing whether the admitted Bob singles and coincidence budget reproduce the observed performance trends. Here, $N_0$ denotes the equivalent background singles spectral density referred to the Bob receiver input immediately before receiver-side spectral filtering; it is not the detected background rate after receiver loss, detector inefficiency, and dead-time saturation. The correct sifted contributions were taken from the anticorrelated matched-basis channels HV/VH and DA/AD, whereas the corresponding error contributions were taken from HH/VV and DD/AA. Representative detector-resolved coincidence channels under reference and high-background conditions are shown in Supplementary Fig. S4. In Fig. 3(a,d,g), Bob total singles rise from 612,206 cps at $\Delta\lambda$ = 0.2 nm to 12,063,420 cps at $\Delta\lambda$ = 5.0 nm, while the sifted-key rate saturates near 0.8–1.0 nm and the QBER increases from 2.68% to 6.32%. In Fig. 3(b,e,h), Bob total singles remain nearly constant, yet the QBER rises from 2.10% at $T_{\mathrm{win}}$ = 50 ps to 17.2% at $T_{\mathrm{win}}$ = 6000 ps and reaches 11.4% already at $T_{\mathrm{win}}$ = 3000 ps, showing that widening the accepted temporal width primarily increases the accepted accidental contribution. In Fig. 3(c,f,i), Bob total singles rise from 318,077 cps at 0 Mcps/nm to 16,713,400 cps at 112.8 Mcps/nm, whereas, under the present operating conditions, the sifted-key rate changes only weakly over the investigated background-noise density, while the error rate and QBER increase with background-noise density. This behavior indicates that the added background photons are manifested primarily through an increased accidental contribution. Representative polarization-resolved Alice and Bob singles under indoor conditions are provided in Supplementary Fig. S3. The model reproduces the dominant trends across all three sweeps when the background-induced accidental budget is referenced to the observed Alice and Bob singles after detector dead-time saturation, which is the same reference plane used in the coincidence analysis. Supplementary Fig. S5 further shows the calculated rate-level decomposition of true, source-originated accidental, and background-induced accidental coincidence contributions, while Supplementary Fig. S6 summarizes the corresponding rate-level multi-click-rejection correction. Because the absolute $N_0$ tolerance depends on the available true-coincidence budget, which changes with source brightness and channel loss, the following design map uses the normalized background-to-signal density ratio, $\rho = N_0/S_0$, to compare receiver operating regimes.

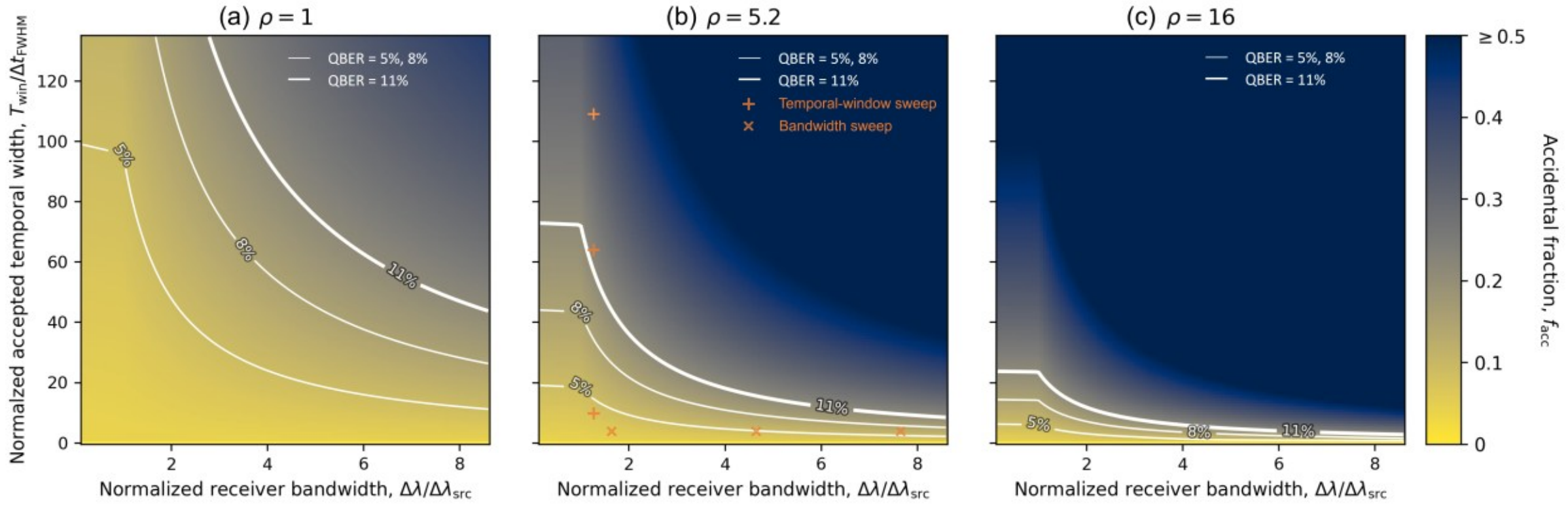

Figure 4. Accidental-fraction landscape with QBER contours under different background-to-signal conditions. Color maps show the accidental fraction, $f_{\text{acc}} = R_{\text{acc}}/(R_{\text{true}} + R_{\text{acc}})$, as a function of normalized receiver bandwidth, $\Delta\lambda/\Delta\lambda_{\text{src}}$, and normalized accepted temporal width, $T_{\text{win}}/\Delta t_{\text{FWHM}}$, for $\rho$ = 1.0, 5.2, and 16.0, where $R_{\text{acc}} = R_{\text{acc,src}} + R_{\text{acc,bg}}$ and $\Delta t_{\text{FWHM}}$ is the measured detector-plane main coincidence-peak FWHM. White contours indicate QBER levels of 5%, 8%, and 11%, with the 11% contour emphasized. The cross markers and plus markers in the $\rho$ = 5.2 panel indicate, respectively, three representative conditions selected from the bandwidth sweep and three from the temporal-window sweep in Fig. 3.

## 2.4 Coupled bandwidth-window landscape of accidental fraction and QBER

Figure 4 provides a system-normalized two-dimensional view of how receiver bandwidth and accepted temporal width jointly determine accidental fraction and QBER. The color scale represents the accidental fraction, $f_{\text{acc}} = R_{\text{acc}}/(R_{\text{true}} + R_{\text{acc}})$, where $R_{\text{acc}} = R_{\text{acc,src}} + R_{\text{acc,bg}}$, which quantifies the contamination of the sifted stream by source-originated and background-induced accidentals, while the overlaid contours denote representative QBER levels of 5%, 8%, and 11%. The map is shown for three representative background-to-signal conditions, $\rho = N_0/S_0$ = 1.0, 5.2, and 16.0, where $N_0$ denotes the background singles spectral density at the Bob input and $S_0$ denotes the signal spectral-density scale at the same reference plane. The middle panel, $\rho$ = 5.2, anchors the design landscape to the representative validation condition used in Fig. 3. Thus, the middle panel provides the direct bridge between the experimentally validated one-dimensional sweeps in Fig. 3 and the two-dimensional receiver-design landscape in Fig. 4. The horizontal axis is the receiver bandwidth normalized to the source bandwidth, $\Delta\lambda/\Delta\lambda_{\mathrm{src}}$, and the vertical axis is the accepted temporal width normalized by the measured detector-plane main coincidence-peak FWHM, $T_{\mathrm{win}}/\Delta t_{\mathrm{FWHM}}$. The × and + markers in the $\rho$ = 5.2 panel indicate representative bandwidth- and temporal-window-sweep conditions from Fig. 3, respectively. To avoid visual crowding, three representative settings were selected from each sweep to span the corresponding one-dimensional trajectory and were projected onto this normalized receiver-parameter plane.

Three features are apparent. First, the accidental fraction increases most strongly toward the upper-right region of the map, where both the normalized receiver bandwidth and normalized temporal acceptance are large. Second, regions of elevated QBER coincide with regions of large accidental fraction, consistent with the rate-model relation between accidental fraction and QBER. Third, increasing ρ compresses the low-QBER operating region toward shorter normalized temporal windows and, at sufficiently high background-to-signal ratio, toward narrower receiver bandwidths. Near the source-matched regime, $\Delta\lambda/\Delta\lambda_{\mathrm{src}}$ = 1, the bandwidth margin remains comparatively broad because the true-coincidence rate is already close to saturation, whereas the normalized temporal-window margin contracts progressively as the background-to-signal ratio increases. The $\rho$ = 5.2 panel anchors this trend to the representative validation condition of Fig. 3, while the $\rho$ = 16.0 panel illustrates a high-background regime in which the 11% QBER contour reaches approximately $T_{\mathrm{win}}/\Delta t_{\mathrm{FWHM}} \approx 23.2$, corresponding to $T_{\mathrm{win}} \approx 1.2$ ns for the present system. Thus, once the receiver bandwidth is approximately matched to the source spectrum, temporal filtering becomes the more sensitive control parameter. The map therefore compactly visualizes how bandwidth and normalized temporal acceptance should be co-varied as the background-to-signal ratio changes.

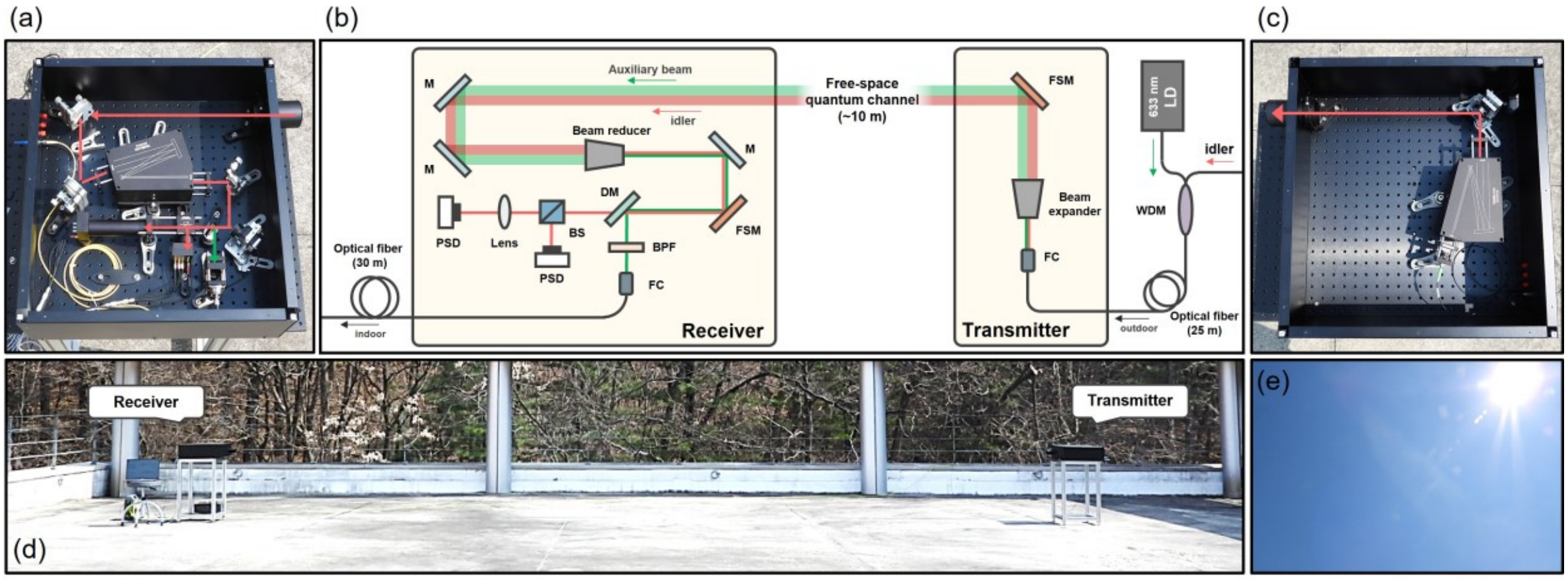

Figure 5. Outdoor free-space demonstration. (a) Photograph of the rooftop receiver optical setup. (b) Schematic of the rooftop experiment. Relative to the indoor validation setup, the indoor quantum-channel section containing the ASE-noise injection stage and 99:1 coupler is removed and replaced by the actual rooftop free-space transmitter and receiver modules. The 1550 nm idler photon is sent to the rooftop transmitter through a 25 m single-mode fiber after combination with a 633 nm auxiliary beam by a 633/1550 WDM. The beam propagates over a 10 m free-space channel, after which the received quantum channel is returned through approximately 30 m of single-mode fiber to the indoor receiver line. Active alignment is implemented using FSMs. At the receiver, a DM separates the 1550 nm quantum channel from the 633 nm auxiliary-alignment beam. In the 1550 nm branch, a 12 nm BPF suppresses residual 633 nm light before fiber coupling, while a 200 GHz DWDM filter in the indoor receiver line provides the QKD spectral filtering. The 633 nm branch is directed to PSD1 and PSD2, with a 125 mm focal-length lens placed before PSD2. (c) Photograph of the rooftop transmitter optical setup. (d) Wide-field photograph of the full rooftop transmitter-receiver arrangement over a 10 m link. (e) Photograph of the cloud-free daytime sky during the measurement period. Abbreviations: WDM, wavelength-division multiplexer; FSM, fast steering mirror; M, mirror; DM, dichroic mirror; BPF, band-pass filter; PSD, position-sensitive detector.

## 2.5 Rooftop free-space demonstration under solar background

The rooftop experiment demonstrates operation of the receiver under real daylight coupling and active-alignment conditions. The field deployment is shown in Fig. 5, where the indoor emulated quantum-channel section was replaced by the actual rooftop transmitter and receiver modules over a 10 m free-space link, with active alignment used during field operation. Figure 6 summarizes the time-dependent rooftop performance in terms of (a) Bob total singles, (b) sifted-key rate, (c) error rate, (d) QBER, and (e) proxy secure-key rate. Over the measurement interval, the system maintained a mean Bob total singles rate of 85,184 ± 9,283 cps, a mean sifted-key rate of 2,811 ± 366 cps, a mean error rate of 125 ± 22 cps, a mean QBER of 4.43 ± 0.29%, and a mean proxy secure-key rate of 1,336 ± 152 cps. These observables exhibited only modest fluctuations during the tested interval, showing stable operation of the rooftop daylight link at the selected receiver settings. This regime was enabled by the preserved true-coincidence budget of the 10 m link together with restricted background acceptance from compact single-mode receiver coupling and 200 GHz dense wavelength-division-multiplexing (DWDM) filtering. Auxiliary field-related measurements, including daylong solar-background monitoring and long-term dual-FSM coupling stability, are provided in Supplementary Note 5.

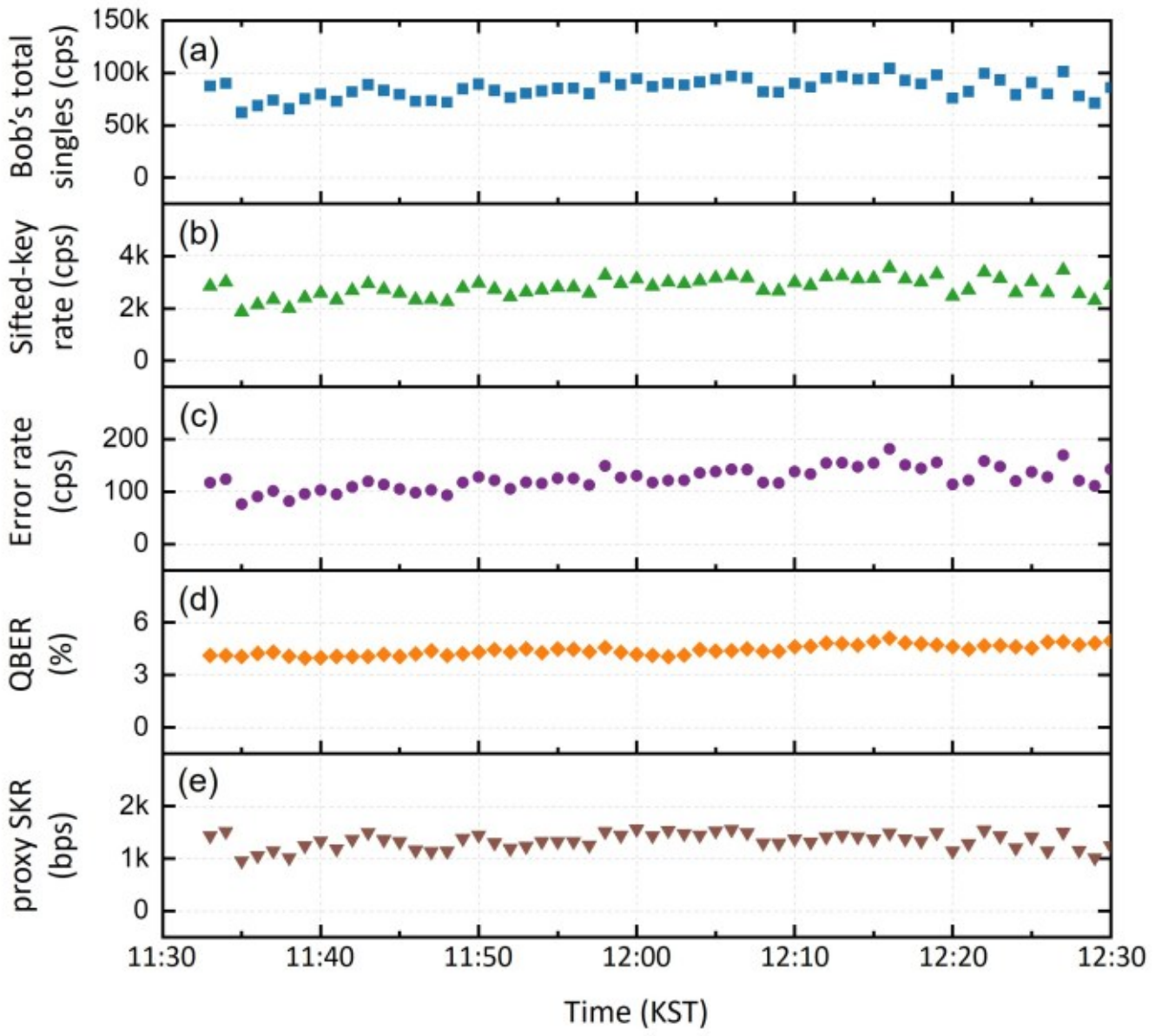

Figure 6. Field performance measured on 23 March 2026 from 11:33 to 12:30 KST under rooftop operation. (a) Bob total singles as a function of time, sampled every 1 min. (b) Sifted-key rate as a function of time. (c) Error rate as a function of time. (d) QBER as a function of time. (e) Proxy secure-key rate (SKR) as a function of time.

## 3. Discussion

The central message of this study is that daylight performance in entanglement-based free-space BBM92 QKD is governed by the accidental-coincidence budget admitted by the receiver. In a pulsed system, that budget is controlled jointly by receiver bandwidth, accepted temporal width, and background-noise density. Treating these quantities as coupled receiver parameters provides a predictive basis for choosing an operating point. The same model, anchored to measured singles rates and reference coincidence components, accounts for the observed trends in Bob singles, sifted-key rate, error rate, and QBER across the three controlled indoor sweeps. This provides an experimentally supported basis for calculating bandwidth–window operating regions at selected QBER levels.

Figures 3 and 4 together sharpen this interpretation. Broadening the receiver bandwidth strongly increases the admitted Bob singles, but the sifted-key rate saturates near the source-matched bandwidth while the QBER continues to rise, showing that additional admitted spectral flux contributes increasingly to accidental contamination rather than useful sifted counts. By contrast, increasing the accepted temporal width leaves Bob total singles nearly unchanged yet increases QBER rapidly because it directly enlarges the temporal overlap probability for uncorrelated detections. The normalized design map in Fig. 4 further shows that this temporal sensitivity becomes more restrictive as the background-to-signal ratio increases. The middle panel, $\rho = 5.2$, anchors the map to the representative validation condition of Fig. 3, whereas the high-background panel, $\rho = 16.0$, illustrates the reduced temporal-window margin more explicitly: near $\Delta\lambda/\Delta\lambda_{\rm src} \approx 1$, the 11% QBER contour occurs at approximately $T_{\mathrm{win}}/\Delta t_{\mathrm{FWHM}} \approx 23.2$, or $T_{\mathrm{win}} \approx 1.2$ ns for the present system. This behavior follows directly from the background-induced accidental scaling with $N_0\, \Delta\lambda\, T_{\mathrm{win}}$.

This interpretation also clarifies the role of receiver filtering. Spectral filtering remains essential because it defines the source-matched operating regime and suppresses out-of-band background photons. Narrowing the receiver bandwidth below the source-matched regime suppresses additional background but also reduces useful signal flux, so the benefit depends on the balance between background rejection and true-coincidence loss. Reducing the accepted temporal width suppresses random overlap without changing Bob singles, but also reduces true-coincidence acceptance when the window becomes comparable to or narrower than the coincidence-peak width. This distinction explains why the measured bandwidth sweep shows limited sifted-key gain beyond the source-matched bandwidth, whereas the temporal-width sweep produces a pronounced QBER increase even when Bob singles remain nearly constant. The calculated rate-level coincidence-component breakdown in Supplementary Fig. S5 provides a transparent view of how the true, source-originated accidental, and background-induced accidental terms evolve across the three sweeps, while Supplementary Fig. S6 summarizes the compact multi-click rejection correction used in the same receiver-level model.

The rooftop experiment provides a complementary demonstration of daylight operation at the selected receiver settings. In the 10 m daylight link, the true-coincidence budget was preserved by the modest channel loss, while compact receiver optics, single-mode-fiber coupling, and telecom-band 200 GHz DWDM filtering restricted the admitted background. Conditions that increase receiver-admitted background relative to the detected true-coincidence signal shift the operating point toward higher accidental fractions, requiring the spectral and temporal acceptance to be reassessed. Changes in receiver aperture can affect both signal collection and background admission, so their effect on QBER depends on the resulting signal and accidental-coincidence rates. Reported daylight background levels and detected noise rates in free-space QKD vary widely with wavelength, receiver aperture, field of view, spectral filtering, and coupling geometry, but prior experiments provide useful numerical context for the admitted background range considered here [13,18-21]. In a 275 m daylight free-space QKD experiment, measured noise counts of 578,565 cps and 289,958 cps were reported under a 1 nm spectral filter for two different spatial-filtering conditions [21], whereas a metropolitan free-space quantum-network experiment reported detected noise rates above 400 kcps in direct sunlight and up to 3.8 times the detected signal rate [18]. In this context, the background range investigated in the present work is practically relevant for daylight free-space QKD while also extending into a demanding stress-test regime. This perspective becomes even more important in longer terrestrial or satellite links, where the detected true-coincidence budget becomes more fragile unless admitted background is tightly constrained at the receiver [13-15,18,19,22].

The present framework is intentionally compact and design-oriented. It reproduces the leading-order behavior of admitted Bob singles, sifted-key rate, error rate, and QBER, but it is not a full microscopic hardware model or a finite-key security proof. The treatment of multi-click rejection is included as a compact rate-level correction rather than as a detector-resolved event taxonomy. The present model uses the measured reference coincidence-peak width and a fixed detector dead time; rate-dependent changes in the coincidence-peak shape and effective detector dead time, as investigated for avalanche photodiodes under strong noise [24], are outside its scope. Similarly, the baseline error term is treated phenomenologically, and polarization-dependent background coupling, atmospheric depolarization, and event-by-event detector dead-time correlations remain natural directions for future refinement. Nevertheless, the agreement between measured and calculated trends across bandwidth, temporal-width, and noise-density sweeps, together with the rooftop demonstration, shows that the receiver-admitted accidental budget provides a practical and experimentally grounded design requirement for daylight entanglement-based free-space BBM92 QKD.

## 4. Conclusions

We developed and tested a receiver-level model anchored to measured singles rates and reference coincidence components for daylight entanglement-based BBM92 QKD. The model reproduces the observed trends in Bob singles, sifted-key rate, error rate, and QBER across controlled sweeps of receiver bandwidth, accepted temporal width, and background-noise density. The calculated design maps identify bandwidth–window combinations compatible with selected QBER levels and show a decreasing temporal-window margin as the background-to-signal ratio increases near source-matched filtering. A complementary rooftop experiment over a 10 m free-space link yielded a mean sifted-key rate of 2,811 cps and a mean QBER of 4.43%. These results provide a quantitative basis for selecting spectral and temporal filtering settings under specified background conditions.

## 5. Methods

### 5.1 Experimental system

The entangled-photon source was a telecom-band polarization-entangled pair source (NuCrypt, EPS-1000-W) operated at a 100 MHz repetition rate. The signal and idler wavelengths were 1555 nm and 1550 nm, respectively. Alice and Bob each used identical polarization-analysis modules consisting of a free-space collimator, a quarter-wave plate (QWP), a half-wave plate (HWP), and a beam splitter (BS). One BS output was analyzed directly by a polarizing beam splitter (PBS) to define the H and V channels, whereas the other passed through an HWP set at 22.5° before a second PBS to define the D and A channels; all PBS outputs were fiber-coupled. Because the source produced an anticorrelated polarization-entangled state, the matched-basis coincidence outcomes HV, VH, DA, and AD were treated as correct sifted events, whereas HH, VV, DD, and AA were treated as error events. Accordingly, $R_{\mathrm{corr}} = R_{\mathrm{HV}} + R_{\mathrm{VH}} + R_{\mathrm{DA}} + R_{\mathrm{AD}}$, $R_{\mathrm{err}} = R_{\mathrm{HH}} + R_{\mathrm{VV}} + R_{\mathrm{DD}} + R_{\mathrm{AA}}$, and $R_{\mathrm{sift}} = R_{\mathrm{corr}} + R_{\mathrm{err}}$. Photons were detected using an SNSPD system (IDQ, ID281) with a nominal specified system detection efficiency of approximately 80% at 1550 nm and a detector dead time of 40 ns. Channel-dependent efficiency and loss variations were accounted for through the experimentally anchored effective transmission factors used in the rate model. Detection events were processed from continuous time-tag streams using a software coincidence window implemented in custom BBM92 analysis software built on the LabVIEW and Python tools provided by Swabian Instruments. $W_{\mathrm{coinc}}$ denotes the Time Tagger coincidence-window setting, such that a detection pair is accepted when $|\Delta t| \leq \mathrm{W}_{\mathrm{coinc}}$. The full accepted temporal width used in the rate model is therefore $T_{\mathrm{win}} = 2W_{\mathrm{coinc}}$. Hereafter, $T_{\mathrm{win}}$ is referred to as the accepted temporal width. Coincidence events were sorted into the corresponding polarization-analysis outcomes for BBM92 evaluation, while multi-click events were counted for diagnostics but excluded from sifted-key and QBER evaluation. Relative channel delays were calibrated from coincidence histograms acquired with the same analysis software for both the indoor and rooftop configurations. A sifted event was accepted only when exactly one click was registered on each side within the coincidence window; groups containing additional clicks were counted and discarded.

**5.2 Controlled indoor background-emulation experiments**

For the controlled indoor experiments, the source output was delivered directly through fiber to the receiver-side testbed rather than through a free-space quantum channel. ASE-based background light was passed through a fixed attenuator and a variable optical attenuator before being combined with the 1550 nm idler channel at Bob using a 99:1 fiber coupler, enabling precise control of the injected background-noise density. The receiver bandwidth was set by a tunable optical filter (Santec, OTF-350). The source spectrum, background spectrum, and filter transmission were measured with an optical spectrum analyzer (Yokogawa, AQ6380). The representative coincidence histogram used to define the timing scale was obtained over 10 s, with a main-peak FWHM of 52 ps. Under the reference low-background condition, the source-originated true and accidental components, 19,893 cps and 834 cps, were obtained from the detector-plane 16-channel summed coincidence histogram. The indoor filter used in this reference measurement had a spectral bandwidth of 0.8 nm and an insertion loss of 3.0 dB, which was treated as constant across the indoor bandwidth sweep. Unless otherwise stated, the indoor baseline system without added background yielded a sifted-key rate of approximately 10,945 cps and a baseline low-background QBER of approximately 2.5%. For the indoor validation experiments in Fig. 3, one control parameter was varied at a time while the others were fixed. For the bandwidth sweep, the background-noise density and accepted temporal width were fixed at 12.8 Mcps/nm and 0.2 ns, respectively. For the accepted-temporal-width sweep, the filter bandwidth and background-noise density were fixed at 0.8 nm and 12.8 Mcps/nm, respectively. For the background-noise-density sweep, the accepted temporal width and filter bandwidth were fixed at 0.2 ns and 0.8 nm, respectively.

**5.3 Coincidence-rate model**

The sifted rate was modeled as $R_{\mathrm{sift}} = R_{\mathrm{true}} + R_{\mathrm{acc}}$, where $R_{\mathrm{acc}} = R_{\mathrm{acc,src}} + R_{\mathrm{acc,bg}}$. In the main text, these quantities denote effective rates after basis sifting and multi-click rejection; the corresponding pre-sifting and pre-rejection components are defined explicitly in Supplementary Note 2. Therefore, no additional factor of 1/2 is applied in the expression for $R_{\mathrm{sift}}$. The true-coincidence and source-originated accidental components were anchored to 19,893 cps and 834 cps, obtained from the detector-plane 16-channel summed coincidence histogram under the low-background reference condition. These quantities were experimentally measured detector-plane anchors. For the model calculation, they were internally converted to source-output-equivalent components only to apply variable receiver-side filtering and loss factors in a common reference plane, and then propagated back to detector-plane rates for comparison with measured observables, as described in Supplementary Note 2. Effective spectral and temporal acceptance factors were set by the receiver filter bandwidth and the coincidence window setting, using the experimentally measured detector-plane main coincidence-peak FWHM of 52 ps as the characteristic timing scale. In the present model, the total accepted temporal width is denoted by $T_{\mathrm{win}}$, with $\mathrm{T_{win}} = 2\mathrm{W_{coinc}}$, where $\mathrm{W_{coinc}}$ is the Time Tagger coincidence-window setting, which defines the maximum allowed event-to-event time difference. The source-originated accidental term was decomposed into multipair and residual source-related contributions [26,27]. The background-induced accidental term was modeled from the admitted Bob-side background singles, with the leading dependence $R_{\mathrm{acc,bg}} \propto N_0 \cdot \Delta\lambda \cdot T_{\mathrm{win}}$ [28]. In the implemented model, the pre-sifting background-induced accidental term was evaluated using the observed detector-plane singles after non-paralyzable dead-time response [29], $C_{\mathrm{acc,bg}} \approx S_{\mathrm{A,tot}}^{\mathrm{obs}} \cdot S_{\mathrm{B,bg}}^{\mathrm{obs}} \cdot T_{\mathrm{win}}$. Basis sifting and multi-click rejection give $R_{\mathrm{acc,bg}} = P_{\mathrm{keep}} p_{\mathrm{match}} C_{\mathrm{acc,bg}}$. The observed Bob background singles were obtained by applying a common dead-time saturation factor determined from the total linear Bob singles rate, including signal, background, and dark counts. This treatment ensures that the accidental budget is referenced to the experimentally measured singles entering the coincidence analysis, rather than to ideal unsaturated photon-arrival rates. Bob total singles were modeled from the admitted signal, background, and dark counts at Bob together with a non-paralyzable dead-time correction using the detector dead time of 40 ns. The experimental multi-click discard rule was approximated in the model using the rate-level survival factor defined in Supplementary Note 2. From these quantities, we obtained the model predictions for the Bob total singles, sifted-key rate, error rate, and QBER compared in Fig. 3. A fuller statement of the quantitative framework is provided in Supplementary Note 2. The background-to-signal ratio used for the middle panel of Fig. 4, $\rho = 5.2$, was derived from the Bob total singles measured with and without added background under the representative validation condition of Fig. 3, after applying the same four-channel non-paralyzable dead-time correction used in the rate model. The representative sweep conditions overlaid in the middle panel of Fig. 4 were obtained by projecting selected Fig. 3 bandwidth- and temporal-window-sweep settings onto the normalized coordinates $\Delta\lambda/\Delta\lambda_{\mathrm{src}}$ and $\mathrm{T_{win}}/\Delta \mathrm{t_{FWHM}}$.

### 5.4 QBER and performance metrics

The QBER was modeled as $Q = \left[E_0 \cdot R_{\mathrm{true}} + (1/2) \cdot \left(R_{\mathrm{acc,src}} + R_{\mathrm{acc,bg}}\right)\right] / R_{\mathrm{sift}}$, where $\mathrm{E_0}$ denotes the intrinsic error probability of true coincidences, and $R_{\mathrm{acc}} = R_{\mathrm{acc,src}} + R_{\mathrm{acc,bg}}$. The measured reference QBER, $Q_0 \approx 2.5\%$, was obtained without added external background and includes residual source-originated accidental coincidences; it is therefore distinguished from $E_0$. The sifted rate was written as $R_{\text{sift}} = R_{\text{true}} + R_{\text{acc}}$, the error rate as $R_{\text{err}} = Q \cdot R_{\text{sift}}$, and the correct-key rate as $R_{\text{corr}} = (1 - Q) \cdot R_{\text{sift}}$. For the rooftop field measurement in Fig. 6, we additionally report a proxy secure-key rate, $R_{\text{SKR}}^{\text{proxy}} = R_{\text{sift}}\,[1 - 2 \cdot h_2(Q)]$, where $h_2$ is the binary entropy, used here only as an operational benchmark rather than as a finite-key security proof [30,31]. Full expressions and implementation details are provided in Supplementary Note 2.

### 5.5 Outdoor field measurement

For the rooftop field measurements in Fig. 6, the indoor noise-injection coupler was removed and the source output was sent to the rooftop transmitter through approximately 25 m of single-mode fiber; after free-space reception, the quantum channel was returned through approximately 30 m of single-mode fiber to the indoor receiver line. Active alignment was implemented using FSMs at the outdoor link. At the receiver, a DM separated the 1550 nm quantum channel from the 633 nm auxiliary-alignment beam; the latter was directed to the PSD-based stabilization path. In the 1550 nm branch, a 12 nm BPF was inserted before fiber coupling to suppress residual 633 nm laser light not fully rejected by the DM. The fiber-coupled 1550 nm quantum channel was then passed through a commercial 200 GHz dense wavelength-division-multiplexing (DWDM) filter in the indoor receiver line, which provided the spectral filtering used for the outdoor QKD measurement and was chosen for its telecom compatibility and comparatively low insertion loss. The Time Tagger coincidence-window setting was $W_{\text{coinc}}$ = 100 ps, corresponding to a total accepted temporal width $T_{\text{win}}$ = 200 ps, to allow margin against outdoor delay drift. Data were acquired on 23 March 2026 from 11:33 to 12:30 with 1 min sampling; the reported mean ± standard deviation values for Bob total singles, sifted-key rate, error rate, QBER, and proxy secure-key rate were calculated over this field time series.

## Data availability

The data that support the findings of this study are available from the corresponding author upon reasonable request.

## Code availability

The custom analysis code used to perform the receiver-level rate-equation calculations and to generate the model curves and design maps in Figs. 3 and 4 and Supplementary Figs. S5 and S6 is available from the corresponding author upon reasonable request.

## Acknowledgements

This study was supported by the Challengeable Future Defense Technology Research and Development program through the Agency for Defense Development (ADD) funded by the Defense Acquisition Program Administration (DAPA) in 2026 (Grant No. 915039201).

## CRediT authorship contribution statement

Jiyoung Moon: Investigation; Data curation; Formal analysis; Methodology; Writing – original draft; Writing – review & editing. Yonggi Jo: Methodology; Validation; Writing – review & editing. Zaeill Kim: Resources; Writing – review & editing. Yong Sup Ihn: Funding acquisition; Resources; Writing – review & editing. Nam Hun Park: Conceptualization; Formal analysis; Investigation; Methodology; Supervision; Writing – review & editing.

## Declaration of competing interest

The authors declare no financial or non-financial competing interests.

**Supplementary Information** for Time-spectral control of accidental coincidences in daylight entanglement-based free-space QKD

This Supplementary Information provides additional background and receiver-filter characterization, a detailed description of the receiver-level rate-equation framework, representative detector-resolved indoor measurements and experimentally anchored model-input parameters, calculated rate-level analyses, and auxiliary field-related measurements supporting the main text.

**Supplementary Note 1. Background and filter characterization**

This note summarizes the experimentally measured spectral and temporal quantities used as inputs to the main-text analysis. It includes the wide-range ASE background spectrum and the measured transmission spectra of the receiver filters.

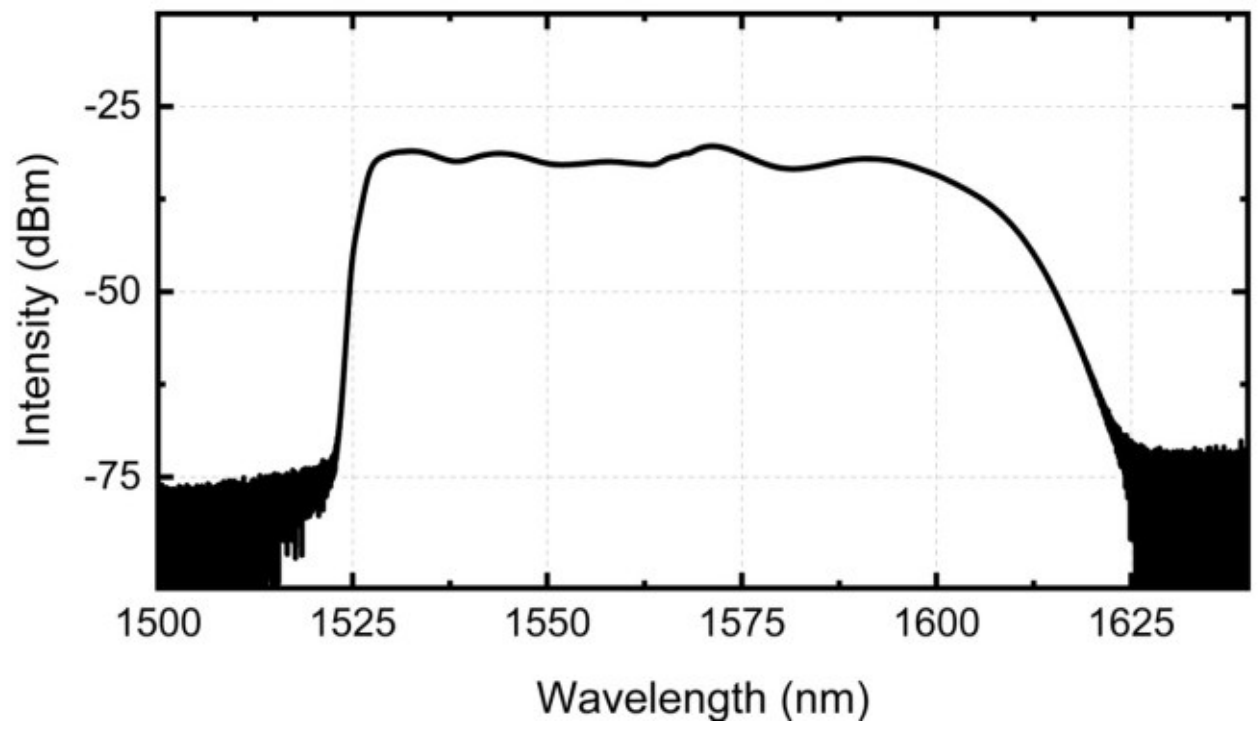

Supplementary Fig. S1. Measured wide-range spectrum of the ASE background source used in the indoor experiments. The spectrum confirms broadband emission over the wavelength range relevant to the receiver-filter study and supports the use of the ASE source as a controllable emulator of background photons in the telecom band.

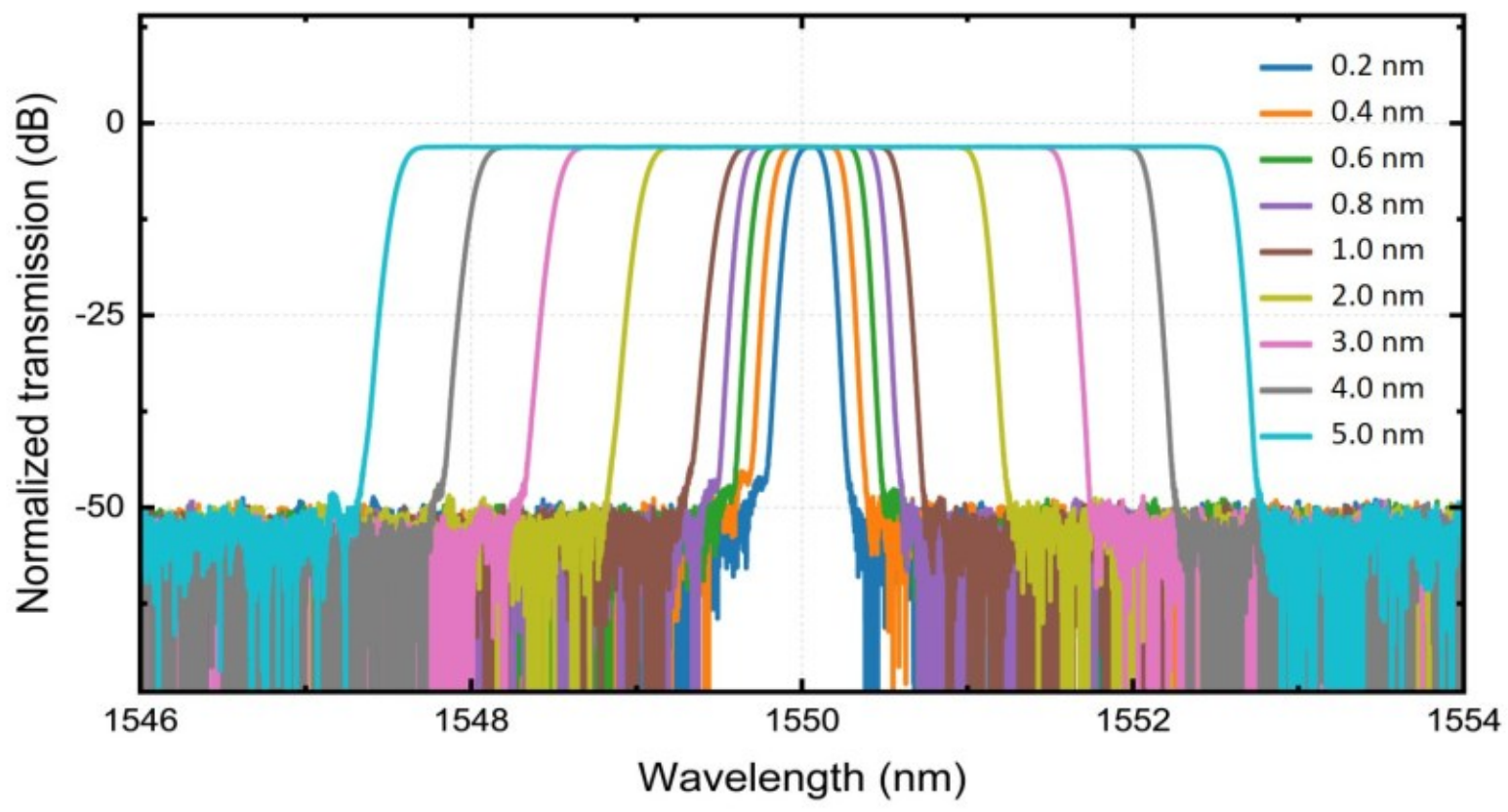

Supplementary Fig. S2. Measured transmission spectra of the receiver filters used in the bandwidth-dependent study. The curves provide the measured center wavelengths, effective passbands, and relative transmission profiles used to determine the effective receiver passbands and to support the bandwidth parameterization in the quantitative framework.

**Supplementary Note 2. Quantitative framework and representative parameters**

To interpret the indoor measurements and generate the time–spectral maps in Figs. 3 and 4, we used a rate-equation model that converts experimentally specified baseline coincidence and singles rates into predicted sifted-key rate, error rate, and QBER under variable receiver bandwidth, accepted temporal width, and background level. The model was implemented in the calculations used for figure generation, in which the source baseline rates, detector-plane singles baselines, receiver losses, detector efficiency, timing jitter, detector dead time, and background-noise density were supplied as experimentally anchored inputs.

For clarity, $C_i$ denotes a coincidence component before basis sifting, $R_i^{\text{pre}}$ denotes the corresponding basis-sifted rate before multi-click rejection, and $R_i$ denotes the final post-rejection effective sifted rate. The index *i* represents true coincidences, source-originated accidentals, or background-induced accidentals. Background-induced quantities are denoted by the subscript bg throughout.

Unless otherwise stated, $C_i$ and $S_X^{\text{obs}}$ denote detector-plane quantities, $N_0$ and $S_0$ are referred to the Bob receiver input immediately before receiver-side spectral filtering, and the superscript (src) denotes source-output-equivalent quantities.

Key notation

| symbol | Meaning |
|---|---|
| $\boldsymbol{N_0}$ | Equivalent Bob-input background spectral density |
| $\boldsymbol{\Delta\lambda}$ | Receiver bandwidth |
| $\boldsymbol{W}_{\text{coinc}}$ | Time Tagger coincidence window setting |
| $\boldsymbol{T}_{\text{win}}$ | Total accepted temporal width, $T_{\text{win}} = 2W_{\text{coinc}}$ |
| $\boldsymbol{S_X^{\text{lin}}}$ | Linear detector-plane singles rate before dead-time saturation |
| $\boldsymbol{S_X^{\text{obs}}}$ | Observed detector-plane singles rate after dead-time saturation |
| $\boldsymbol{C_i}$ | Coincidence component before basis sifting |
| $\boldsymbol{R_i^{\text{pre}}}$ | Basis-sifted rate before multi-click rejection |
| $\boldsymbol{R_i}$ | Post-rejection effective sifted rate |
| $\boldsymbol{P}_{\text{keep}}$ | Exact-1&1 survival probability |
| $Q$ | Quantum bit error rate |

2.1 Baseline coincidence and singles rates

The model was anchored to detector-plane coincidence and singles measurements acquired under the low-background reference condition. The reference true-coincidence component $C_{\mathrm{true},0}$ and the source-originated accidental-coincidence component $C_{\mathrm{acc,src},0}$ were obtained from the 16-channel summed coincidence histogram, and the Alice and Bob signal singles were taken from the corresponding detector-plane singles measurements. These detector-plane quantities provide the experimental anchor for comparison with the measured QKD observables. For the calculation, the detector-plane anchors were internally converted to source-output-equivalent components only to apply variable receiver-side filtering and loss factors in a common reference plane. The predicted rates were then mapped back to the detector plane for comparison with the measured Bob singles, sifted-key rate, error rate, and QBER.

The linear detected singles at Alice were written as

$$S_{\mathrm{A,\,tot}}^{\mathrm{lin}} = S_{\mathrm{A,\,sig}}^{\mathrm{lin}} + S_{\mathrm{A,dark}},$$

and the corresponding baseline linear detected singles at Bob were written as

$$S_{\mathrm{B,\,tot}}^{\mathrm{lin}} = S_{\mathrm{B,\,sig}}^{\mathrm{lin}} + S_{\mathrm{B,dark}}.$$

These linear singles represent the detector-plane rates before the detector dead-time response. The observed singles used in the accidental-coincidence calculation were obtained after applying the dead-time saturation model described below.

## 2.2 Detector dead-time saturation

Detector dead-time saturation was treated using a non-paralyzable dead-time model. Because both Alice and Bob employ four-channel polarization-analysis modules, the total singles on each side were approximated as being distributed over $N_{\mathrm{det}} = 4$ detector channels.

For side $X$ = A, B, the observed total singles were therefore written as

$$S_{X,\,\mathrm{tot}}^{\mathrm{obs}} = \frac{S_{X,\,\mathrm{tot}}^{\mathrm{lin}}}{1 + (S_{X,\,\mathrm{tot}}^{\mathrm{lin}} \tau_{\mathrm{d}} / N_{\mathrm{det}})},$$

where $\tau_{\mathrm{d}}$ is the detector dead time. This expression approximates the channel-wise non-paralyzable response after the four-way polarization splitting.

The same dead-time transformation was applied to the relevant singles components on both sides before they were used in accidental-coincidence estimates. In particular, the Alice total singles and the Bob background singles entering the background-induced accidental term below are observed rates after detector dead-time saturation, not ideal unsaturated photon-arrival rates. We used a non-paralyzable dead-time response, which is standard for photon-counting rate corrections [S1].

2.3 Effective source bandwidth and spectral truncation

The effective source bandwidth was taken as the source bandwidth after internal filtering, denoted by $\Delta\lambda_{\mathrm{src}}$. In the present model, $\Delta\lambda_{\mathrm{src}}$ was set to 0.64 nm, using the measured source spectral FWHM as an approximation to the effective bandwidth (main-text Fig. 2(d) and Supplementary Table S1). The collected signal fraction under a receiver filter of bandwidth $\Delta\lambda$ was modeled as

$$f_{\mathrm{sig}}(\Delta\lambda) = \min\left(1, \frac{\Delta\lambda}{\Delta\lambda_{\mathrm{src}}}\right),$$

so that the detected true-coincidence component scales linearly with receiver bandwidth below the source bandwidth and saturates once the full source spectrum is transmitted. In the present effective model, the same bandwidth-limiting factor was applied to the source-origin accidental terms, with the multipair contribution approximated to scale as $\left[f_{\mathrm{sig}}(\Delta\lambda)\right]^2$.

2.4 Pulsed-source decomposition of source-originated accidentals

The source was treated as a pulsed pair source with repetition rate $f_{\mathrm{rep}}$. From the source-output true-coincidence rate, the mean pair number per pulse was estimated as

$$\mu = \frac{C_{\mathrm{true},0}^{(\mathrm{src})}}{f_{\mathrm{rep}}}.$$

The source-originated accidental component was then decomposed into a multipair contribution and a residual contribution. The multipair accidental rate at the source output was written as

$$C_{\mathrm{acc,mp}}^{(\mathrm{src})} = f_{\mathrm{rep}}\mu^2,$$

and the residual source accidental rate was taken as

$$C_{\mathrm{acc,other}}^{(\mathrm{src})} = C_{\mathrm{acc,src},0}^{(\mathrm{src})} - C_{\mathrm{acc,mp}}^{(\mathrm{src})},$$

clipped to zero if necessary. These components were then mapped to the detector plane using the modeled coincidence detection efficiency. This treatment is consistent with standard multipair scaling arguments for spontaneous parametric sources and with prior analyses of their impact on entanglement visibility and QKD performance [S2,S3].

2.5 Temporal acceptance from coincidence-window duration and timing jitter

The coincidence-window acceptance was modeled using a Gaussian timing-jitter response with standard deviation

$$\sigma_t = \frac{\Delta t_{\mathrm{FWHM}}}{2.355}.$$

$W_{\mathrm{coinc}}$ denotes the Time Tagger coincidence-window setting, i.e., the maximum allowed time difference between two detection events. The total accepted temporal width used in the rate model is $T_{\mathrm{win}} = 2W_{\mathrm{coinc}}$. After this definition, $T_{\mathrm{win}}$ is used throughout the rate equation.

For a total accepted temporal width $T_{\mathrm{win}}$, the temporal collection fraction was

$$f_t(T_{\mathrm{win}}) = \mathrm{erf}\left[\frac{T_{\mathrm{win}} / 2}{\sqrt{2}\,\sigma_t}\right].$$

The true-coincidence rate and source-originated accidental components were scaled by this temporal acceptance factor relative to the reference window used in the model. In this way, narrowing the accepted temporal width suppresses both the accepted true-coincidence events and the temporally overlapping source-originated accidental contribution.

2.6 Background-noise singles and noise-induced accidentals

Background noise was introduced only on the Bob side. The model uses $N_0$ as the equivalent background singles spectral density at the Bob input. For high-background settings, $N_0$ was inferred from below-saturation measurements with an additional ND filter and corrected for the measured ND-filter and receiver/filter attenuation. For receiver bandwidth $\Delta\lambda$, the background singles admitted at the Bob input were

$$S_{\mathrm{B,bg}}^{\mathrm{in}} = N_0 \Delta\lambda.$$

The corresponding linear detector-plane Bob background singles before dead-time saturation were

$$S_{\mathrm{B,\,bg}}^{\mathrm{lin}} = \eta_{\mathrm{B}} N_0 \Delta\lambda,$$

where $\eta_{\mathrm{B}}$ is the total Bob-side detection efficiency including receiver optics, filter transmission, and detector efficiency. The observed Bob background singles were estimated by applying a common dead-time saturation factor determined from the total linear Bob singles rate, including signal, background, and dark counts:

$$S_{\mathrm{B,\,bg}}^{\mathrm{obs}} = \frac{S_{\mathrm{B,\,bg}}^{\mathrm{lin}}}{1 + (S_{\mathrm{B,tot}}^{\mathrm{lin}} \tau_{\mathrm{d}} / N_{\mathrm{det}})}.$$

where $S_{\mathrm{B,tot}}^{\mathrm{lin}} = S_{\mathrm{B,sig}}^{\mathrm{lin}} + S_{\mathrm{B,bg}}^{\mathrm{lin}} + S_{\mathrm{B,dark}}$ is evaluated at the corresponding receiver bandwidth and background level.

For the calculations used in Fig. 3, the background-induced accidental coincidence rate before basis sifting was written as

$$C_{\mathrm{acc,bg}} = S_{\mathrm{A,tot}}^{\mathrm{obs}} S_{\mathrm{B,bg}}^{\mathrm{obs}} T_{\mathrm{win}}.$$

where $S_{\mathrm{A,tot}}^{\mathrm{obs}}$ is the observed total Alice singles rate after detector dead-time saturation, $S_{\mathrm{B,bg}}^{\mathrm{obs}}$ is the observed Bob background singles rate after detector dead-time saturation, and $T_{\mathrm{win}} = 2W_{\mathrm{coinc}}$ is the total accepted temporal width corresponding to the Time Tagger coincidence-window setting $W_{\mathrm{coinc}}$. Here both singles rates are detector-plane observed rates after dead-time saturation, so the accidental estimate is evaluated at the same reference plane as the measured coincidence observables. This product-of-singles form follows the standard accidental-coincidence approximation for statistically independent detection streams [S4]. In this observed-rate formulation, detector dead time is included through the singles rates in the product above. The resulting coincidence components are then subjected to basis sifting and multi-click rejection as defined in Section 2.7.

## 2.7 Basis sifting and multi-click handling

For BBM92, basis sifting was modeled with a matched-basis probability of

$$p_{\mathrm{match}} = \frac{1}{2},$$

so that each coincidence component before basis sifting was converted into a pre-rejection sifted-rate component as

$$R_i^{\mathrm{pre}} = p_{\mathrm{match}} C_i,$$

where $i \in \{\text{true, acc,src, acc,bg}\}$. The pre-rejection sifted coincidence budget was then written as

$$R_{\text{sift}}^{\text{pre}} = R_{\text{true}}^{\text{pre}} + R_{\text{acc,src}}^{\text{pre}} + R_{\text{acc,bg}}^{\text{pre}}.$$

After multi-click rejection, the total accidental rate and the final sifted rate are defined as $R_{\text{acc}} = R_{\text{acc,src}} + R_{\text{acc,bg}}$ and $R_{\text{sift}} = R_{\text{true}} + R_{\text{acc}}$, respectively. Therefore, no additional factor of 1/2 is applied in the final expression for $R_{\text{sift}}$.

Multi-clicks were treated as a rate-level exact-1&1 rejection correction. In the experimental post-processing, events containing more than one click on Alice's side or Bob's side within the accepted coincidence window were counted and discarded. In the model, this invalid-event rejection was represented by a survival probability $P_{\text{keep}}$ applied to the basis-sifted coincidence budget. Threshold-detector models and entanglement-based free-space QKD implementations are discussed in Refs. [S5,S6]. The survival probability used here is a phenomenological approximation to our experimental counted-and-discarded rule and does not constitute a security proof for that rule.

The mean number of additional clicks within the accepted temporal window was approximated as

$$\lambda_{\text{extra}} = \left[\frac{(N_{\text{det}} - 1)}{N_{\text{det}}}\right] \left(S_{\text{A,tot}}^{\text{obs}} + S_{\text{B,tot}}^{\text{obs}}\right) T_{\text{win}},$$

where $N_{\text{det}} = 4$, $S_{\text{A,tot}}^{\text{obs}}$ and $S_{\text{B,tot}}^{\text{obs}}$ are the detector-plane observed total singles rates after dead-time saturation, and $T_{\text{win}} = 2W_{\text{coinc}}$. The exact-1&1 survival probability was

$$P_{\text{keep}} = \mathrm{e}^{-\lambda_{\text{extra}}}.$$

The post-rejection components are obtained as

$$R_i = P_{\text{keep}} R_i^{\text{pre}},$$

where $i \in \{\text{true, acc,src, acc,bg}\}$. Therefore,

$$R_{\text{sift}} = P_{\text{keep}} R_{\text{sift}}^{\text{pre}},$$

and the estimated rejected multi-click rate is

$$R_{\mathrm{mc}} = R_{\mathrm{sift}}^{\mathrm{pre}}\left(1 - P_{\mathrm{keep}}\right) = R_{\mathrm{sift}}^{\mathrm{pre}} - R_{\mathrm{sift}}.$$

This term was used as a compact rate-level correction for invalid-event rejection, rather than as a detector-resolved multi-click taxonomy.

2.8 QBER, sifted-key rate, and error rate

In the following QBER calculation, all rates without the superscript pre denote post-rejection effective sifted rates. Thus, $R_{\mathrm{true}}$, $R_{\mathrm{acc,src}}$, $R_{\mathrm{acc,bg}}$, $R_{\mathrm{acc}}$ and $R_{\mathrm{sift}}$ denote final effective rates after basis sifting and multi-click rejection.

The measured low-background reference QBER is denoted by $Q_0 \approx 2.5\%$. It includes the baseline errors observed without added external background, including residual source-originated accidental coincidences. In the model, $E_0$ denotes the intrinsic optical error probability associated with true coincidences. Under the same reference window and rate-processing convention, these quantities are related by

$$Q_0 = \frac{E_0 R_{\mathrm{true},0} + \frac{1}{2} R_{\mathrm{acc,src},0}}{R_{\mathrm{true},0} + R_{\mathrm{acc,src},0}}.$$

All rates in this reference relation are post-rejection effective sifted rates. The total accidental rate was defined as

$$R_{\mathrm{acc}} = R_{\mathrm{acc,src}} + R_{\mathrm{acc,bg}},$$

and the final sifted rate was written as

$$R_{\mathrm{sift}} = R_{\mathrm{true}} + R_{\mathrm{acc}}.$$

Thus, the QBER was calculated as

$$Q = \frac{E_0 R_{\mathrm{true}} + \frac{1}{2} R_{\mathrm{acc}}}{R_{\mathrm{sift}}},$$

The error and correct-key rates were then calculated as

$$R_{\mathrm{err}} = Q R_{\mathrm{sift}},$$

$$R_{\mathrm{corr}} = (1 - Q) R_{\mathrm{sift}}.$$

Here $R_{\mathrm{err}}$ and $R_{\mathrm{corr}}$ are effective rate-level quantities derived from the modeled QBER. The 50% accidental-error contribution reflects the random polarization outcome expected for uncorrelated accidental coincidences.

A proxy secure-key-rate metric was also evaluated as

$$R_{\mathrm{SKR}}^{\mathrm{proxy}} = R_{\mathrm{sift}}[1 - 2h_2(Q)],$$

with the rate clipped to zero for $Q > 11\%$. This quantity was used only as a practical operating-region indicator and not as a finite-key security proof. The binary-entropy-based asymptotic secure-fraction expression and the corresponding ~11% QBER benchmark follow the standard one-way asymptotic BB84/BBM92 security interpretation [S7,S8].

2.9 Landscape parameterization for Fig. 4

For Fig. 4, the background level was parameterized by the background-to-signal spectral-density ratio

$$\rho = \frac{N_0}{S_0},$$

where $N_0$ and $S_0$ denote the background and signal singles spectral-density scales, respectively, referred to the same Bob-input reference plane. In the main-text Fig. 4, representative maps were evaluated for $\rho$=1.0, 5.2, and 16.0. For each $\rho$, the model computed two-dimensional maps over the ($\Delta\lambda$, $T_{\mathrm{win}}$) plane using the same accidental-budget framework described above.

The representative value $\rho = 5.2$ was derived from the Bob total singles measured with and without added background under the validation condition of Fig. 3, using the same four-channel non-paralyzable dead-time correction as the rate model.

The plotted colour metric was the accidental fraction

$$f_{\mathrm{acc}} = \frac{R_{\mathrm{acc}}}{R_{\mathrm{true}} + R_{\mathrm{acc}}} = \frac{R_{\mathrm{acc,src}} + R_{\mathrm{acc,bg}}}{R_{\mathrm{true}} + R_{\mathrm{acc,src}} + R_{\mathrm{acc,bg}}},$$

where $R_{\mathrm{acc}} = R_{\mathrm{acc,src}} + R_{\mathrm{acc,bg}}$, while the contours represented QBER. This construction makes Fig. 4 a receiver-design map: the colour indicates how much of the sifted stream is contaminated by accidental events, and the contour lines show the corresponding error level.

2.10 Scope and limitations of the model

The framework is intentionally compact and phenomenological. It is designed to capture the dominant receiver-level mechanism governing the measured performance under background illumination: spectral truncation of the signal, temporal acceptance of coincidence events, source-originated accidentals, background-induced accidentals, detector-dead-time saturation, and rate-level invalid-event rejection. It does not include finite-key effects, polarization-dependent background coupling, event-by-event dead-time correlations, or a microscopic source-state model. These effects are outside the scope of the present receiver-level validation.

**Supplementary Note 3. Representative detector-resolved indoor measurements and model inputs**

This note summarizes representative detector-resolved indoor measurements and the principal measured values and experimentally anchored inputs used in the compact receiver-level model.

3.1 Polarization-resolved single-photon count rates under representative indoor conditions

To support the detector-plane singles interpretation used in the main text, Supplementary Fig. S3 presents representative polarization-resolved single-photon count rates for both Alice and Bob under two indoor conditions.

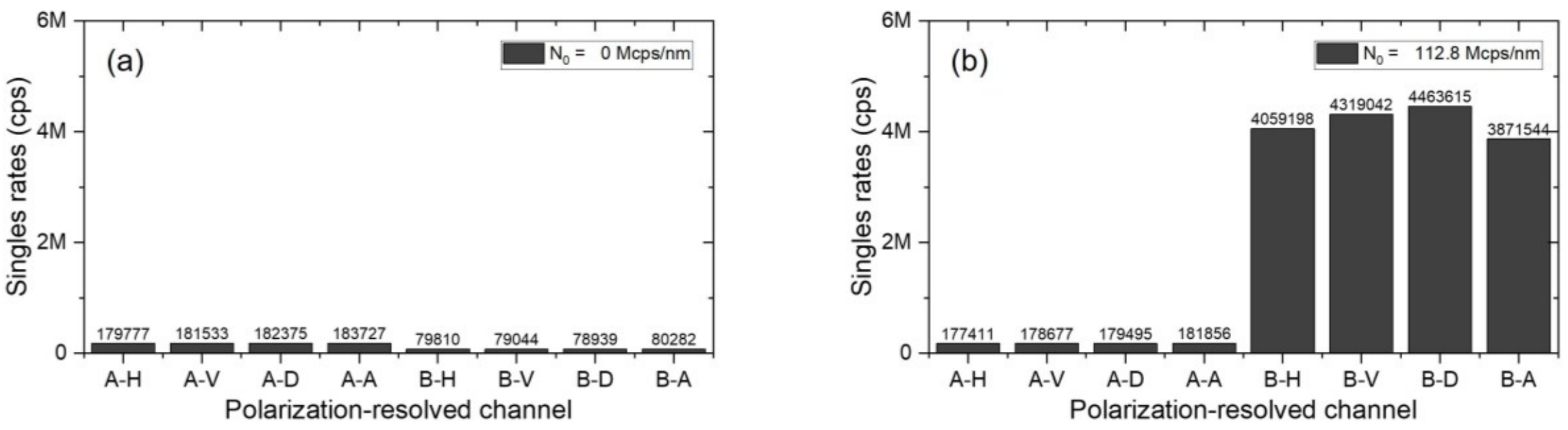

Supplementary Fig. S3. Polarization-resolved singles for the four Alice and Bob analysis channels (H, V, D, A) under two representative indoor conditions: (a) reference condition, defined by $\Delta\lambda$ = 0.8 nm, $W_{\text{coinc}}$ = 0.1 ns ($T_{\text{win}}$ = 0.2 ns), and $N_0$ = 0 Mcps/nm, and (b) high-background condition, defined by $\Delta\lambda$ = 0.8 nm, $W_{\text{coinc}}$ = 0.1 ns ($T_{\text{win}}$ = 0.2 ns), and $N_0$ = 112.8 Mcps/nm. The numerical single-photon count rates are indicated above each bar.

The comparison provides a compact detector-plane view of how the admitted background increases the single-photon count rates across the Bob polarization-analysis channels while the polarization-resolved singles remain comparable among the four analysis channels under the common receiver setting.

3.2 Detector-resolved coincidence channels under representative indoor conditions

To support the detector-resolved BBM92 channel assignment used in the main text, Supplementary Fig. S4 presents two representative bar plots of the eight polarization-resolved coincidence channels (HH, HV, VH, VV, DD, DA, AD, AA). The comparison provides a compact detector-resolved view of how the coincidence structure changes between a nominal low-background operating point and a strongly background-contaminated condition. In the reference condition, the correct sifted contributions (HV, VH, DA, AD) remain larger than the error contributions (HH, VV, DD, AA). Under high background, the relative contribution of the error channels increases, consistent with the accidental-budget interpretation developed in the main text.

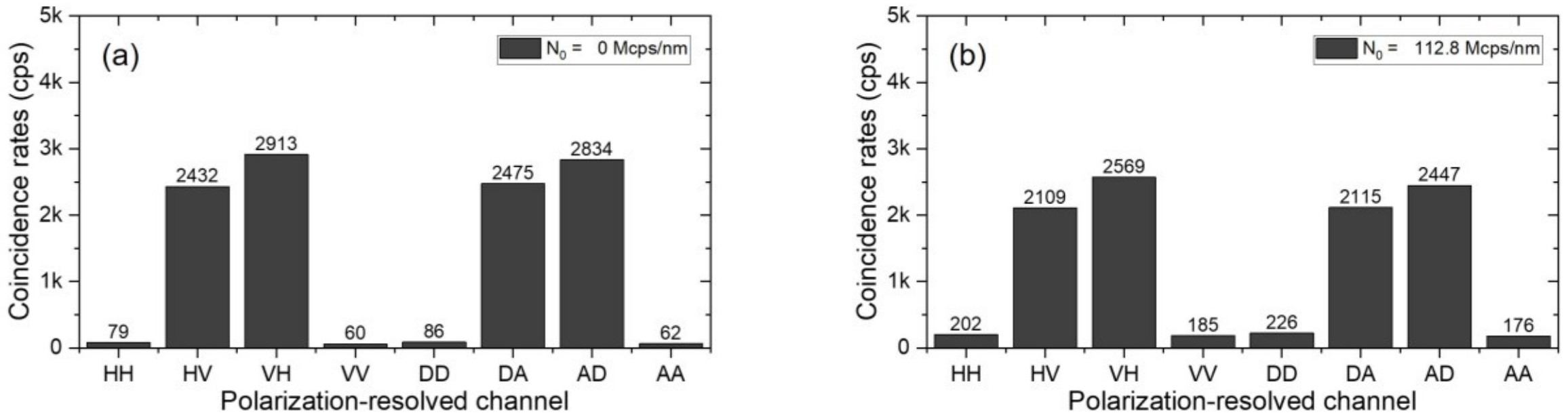

Supplementary Fig. S4. Detector-resolved coincidence rates for the eight polarization combinations (HH, HV, VH, VV, DD, DA, AD, AA) under two representative indoor conditions: (a) reference condition, defined by $\Delta\lambda$ = 0.8 nm, $W_{\text{coinc}}$ = 0.1 ns ($T_{\text{win}}$ = 0.2 ns), and $N_0$ = 0 Mcps/nm, and (b) high-background condition, defined by $\Delta\lambda$ = 0.8 nm, $W_{\text{coinc}}$ = 0.1 ns ($T_{\text{win}}$ = 0.2 ns), and $N_0$ = 112.8 Mcps/nm. The numerical coincidence rates are indicated above each bar.

### 3.3 Principal measured values and model-input parameters

For clarity, Supplementary Table S1 summarizes the principal experimental anchor values and model-input parameters used in the receiver-level rate-equation framework. The coincidence reference values were obtained from the detector-plane 16-channel summed coincidence histogram, and the singles reference values were obtained from the corresponding detector-plane singles measurements under the low-background reference condition. Because the detector efficiencies and channel losses were not perfectly identical across the polarization-resolved detection channels, the fiber polarization controllers were adjusted to balance the HVDA channel rates before the main measurements. The effective Alice- and Bob-side transmission factors used in the model therefore represent channel-balanced effective values rather than bare component-by-component losses.

Supplementary Table S1. Principal experimental anchor values and model-input parameters used in the receiver-level rate-equation framework. Coincidence and singles reference values are detector-plane measured quantities under the low-background reference condition. Effective transmission factors represent channel-balanced values used in the model.

| Quantity | Symbol | Measured value | Input value used in model | Note |
|---|---|---|---|---|
| Source-originated true coincidence rate under the reference condition | $C_{\mathrm{true},0}$ | 19,893 cps | 19,893 cps | From detector-plane 16-channel summed histogram |
| Source-originated accidental coincidence rate under the reference condition | $C_{\mathrm{acc,src},0}$ | 834 cps | 834 cps | From detector-plane 16-channel summed histogram |
| Main-peak full width at half maximum | $\Delta t_{\mathrm{FWHM}}$ | 52 ps | 52 ps | From detector-plane 16-channel summed histogram |
| Reference Alice total singles | $S_{\mathrm{A,tot,ref}}$ | 727,412 cps | 727,412 cps | Detector plane reference singles |
| Reference Bob total singles | $S_{\mathrm{B,tot,ref}}$ | 318,077 cps | 318,077 cps | Detector plane reference singles |
| Effective Alice-side transmission from the reference plane to the detector plane | $\eta_{\mathrm{A,eff}}$ | 0.44 | 0.44 | Channel-balanced effective value |
| Effective Bob-side transmission from the reference plane to the detector plane | $\eta_{\mathrm{B,eff}}$ | 0.20 | 0.20 | Channel-balanced effective value |
| Baseline low-background QBER | $Q_0$ | 2.5% | 2.5% (reference QBER) | Measured without added external background |

| Quantity | Symbol | Measured value | Input value used in model | Note |
|---|---|---|---|---|
| Intrinsic true-coincidence error probability | $E_0$ | — | 0.5% | Distinct from the measured reference QBER |
| Detector dead time | $\tau_\mathrm{d}$ | 40 ns | 40 ns | Measured value |
| Receiver filter bandwidth | $\Delta\lambda$ | Variable | Variable | Measured passband |
| Coincidence-window setting | $W_\mathrm{coinc}$ | Variable | Variable | Time Tagger setting |
| Total accepted temporal width | $T_\mathrm{win}$ | $2W_\mathrm{coinc}$ | $2W_\mathrm{coinc}$ | Model quantity |
| Background-noise density | $N_0$ | Variable | Variable | Equivalent Bob-input density from below-saturation ND-filter measurement; corrected for attenuation |
| Indoor filter bandwidth used for the reference characterization | $\Delta\lambda_\mathrm{indoor}$ | 0.8 nm | 0.8 nm | Measured passband |
| Indoor filter insertion loss | $\mathrm{IL}_{\Delta\lambda,\mathrm{indoor}}$ | 3.0 dB | 3.0 dB | Measured value |
| Source repetition rate | $f_\mathrm{rep}$ | 100 MHz | 100 MHz | Source operating setting reported in Methods |
| Detector channels per receiver | $N_\mathrm{det}$ | 4 | 4 | Four-channel approximation in Note 2.2 |
| Reference accepted temporal width | $T_\mathrm{win,ref}$ | 0.2 ns | 0.2 ns | Reference setting in Figs. S3 and S4 |

| Quantity | Symbol | Measured value | Input value used in model | Note |
|---|---|---|---|---|
| Measured source spectral FWHM | $\Delta\lambda_{\mathrm{FWHM,src}}$ | 0.64 nm | 0.64 nm | Measured spectrum in main-text Fig. 2(d) |

**Supplementary Note 4. Auxiliary calculated rate-level analysis**

4.1 Calculated rate-level coincidence-component breakdown

To support the interpretation of the measured trends in Fig. 3, we calculated the rate-level decomposition of the basis-sifted coincidence budget into true, source-originated accidental, and background-induced accidental contributions. The pre-rejection sifted coincidence budget was written as

$$R_{\mathrm{sift}}^{\mathrm{pre}} = R_{\mathrm{true}}^{\mathrm{pre}} + R_{\mathrm{acc,src}}^{\mathrm{pre}} + R_{\mathrm{acc,bg}}^{\mathrm{pre}}.$$

Here $R_{\mathrm{true}}^{\mathrm{pre}}$, $R_{\mathrm{acc,src}}^{\mathrm{pre}}$, and $R_{\mathrm{acc,bg}}^{\mathrm{pre}}$ denote the calculated true, source-originated accidental, and background-induced accidental coincidence contributions before the compact exact-1&1 multi-click-rejection correction is applied. Supplementary Fig. S5 shows these calculated rate-level components for the receiver-bandwidth, accepted-temporal-width, and background-noise-density sweeps. A logarithmic y-axis is used to display the true and accidental components on the same scale.

This decomposition is a model-inferred rate-level breakdown obtained from the same experimentally anchored parameters used in Fig. 3. The individual components should not be interpreted as independently measured event-by-event labels.

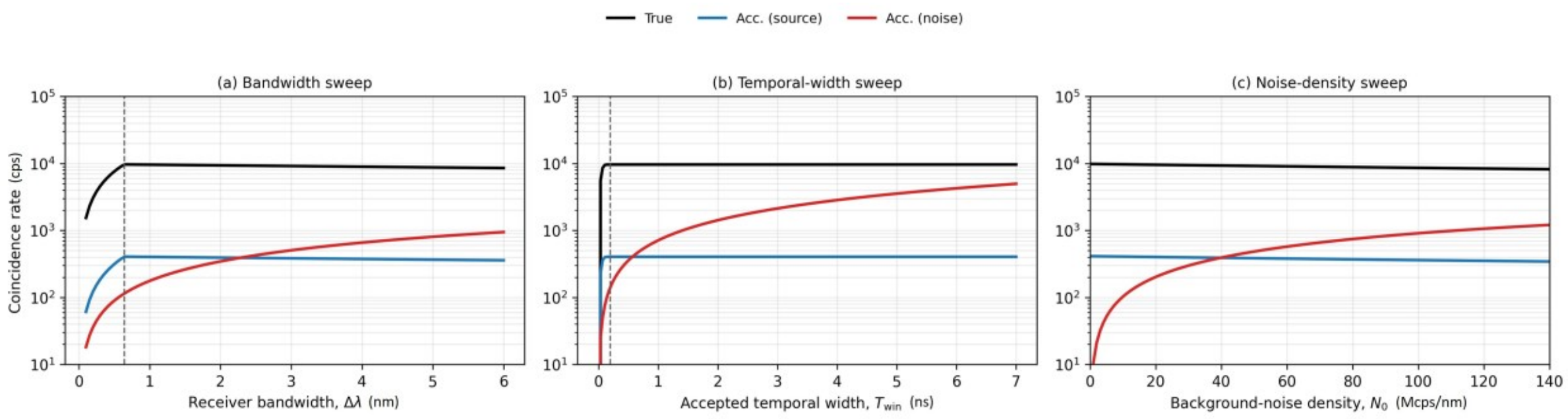

Supplementary Fig. S5. Calculated rate-level coincidence-component breakdown for the three indoor sweeps. The panels show the calculated rates of true coincidences, source-originated accidentals, and background-induced accidentals for the three indoor sweeps: (a) receiver-bandwidth sweep, (b) accepted-temporal-width sweep, and (c) background-noise-density sweep. The plotted quantities are evaluated before the compact exact-1&1 multi-click-rejection correction is applied. The logarithmic y-axis highlights the relative magnitudes and trends of the true and accidental components over the full sweep range. The decomposition is calculated using the same experimentally anchored model parameters as Fig. 3 and is intended to visualize the rate-level coincidence budget rather than to provide independently measured event-by-event component labels.

## 4.2 Calculated rate-level multi-click-rejection correction

This section presents the calculated rate-level multi-click-rejection term used in the receiver-level rate-equation model for the main-text calculations. In the experimental post-processing, coincidence candidates containing more than one click on Alice's side or Bob's side within the accepted coincidence window were counted and discarded. In the model, the corresponding rejected rate was estimated from the difference between the pre-rejection and post-rejection sifted coincidence budgets,

$$R_{\mathrm{mc}} = R_{\mathrm{sift}}^{\mathrm{pre}} - R_{\mathrm{sift}}.$$

For each sweep, the plotted fraction was evaluated as

$$f_{\mathrm{mc,est}} = \frac{R_{\mathrm{mc}}}{R_{\mathrm{sift}}^{\mathrm{pre}}}.$$

This quantity represents the estimated fraction of basis-sifted coincidence candidates rejected by the counted-and-discarded multi-click policy. It is included only as an auxiliary rate-level correction and is not a detector-resolved multi-click taxonomy.

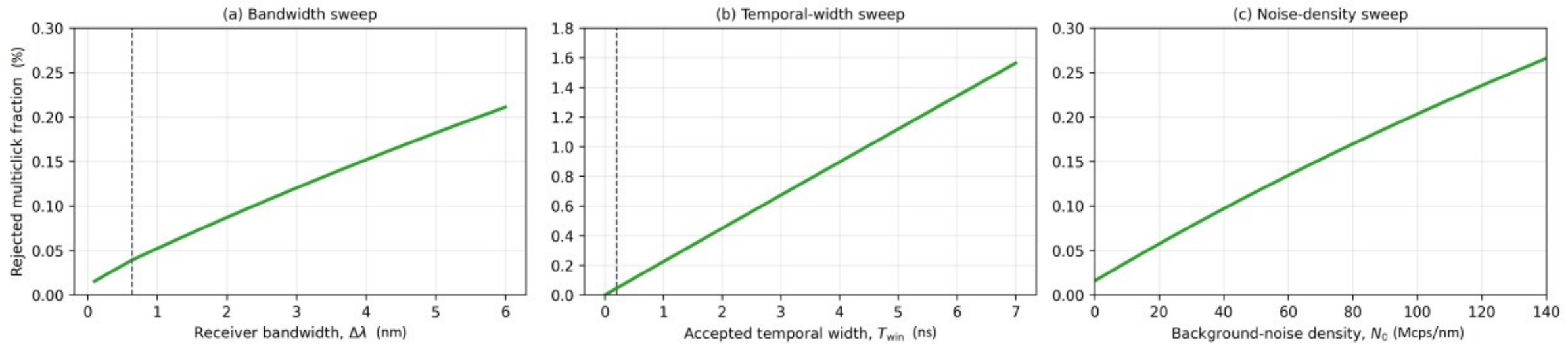

Supplementary Fig. S6. Calculated rate-level multi-click-rejection correction for the three indoor sweeps. The panels show the estimated fraction of basis-sifted coincidence candidates rejected by the counted-and-discarded multi-click policy for the three indoor sweeps: (a) receiver-bandwidth sweep, (b) accepted-temporal-width sweep, and (c) background-noise-density sweep. The plotted quantity is $100 f_{\mathrm{mc,est}}$ (%), where $f_{\mathrm{mc,est}} = R_{\mathrm{mc}}/R_{\mathrm{sift}}^{\mathrm{pre}}$ and $R_{\mathrm{mc}} = R_{\mathrm{sift}}^{\mathrm{pre}} - R_{\mathrm{sift}}$. The calculation provides a compact rate-level estimate of invalid-event rejection.

## Supplementary Note 5. Field-related auxiliary measurements

This note provides auxiliary measurements related to the outdoor free-space experiment, including the daylong solar-background level incident on the receiver optics and the long-term coupling stability obtained under the dual-FSM alignment configuration.

### 5.1 Daylong solar-background measurement at the receiver optics

This section presents a daylong measurement of solar-background singles collected at the receiver optics without the polarization-analysis stage and without receiver filtering. The purpose of this measurement is to provide a reference for the magnitude and temporal variation of the daytime background relevant to the outdoor free-space environment.

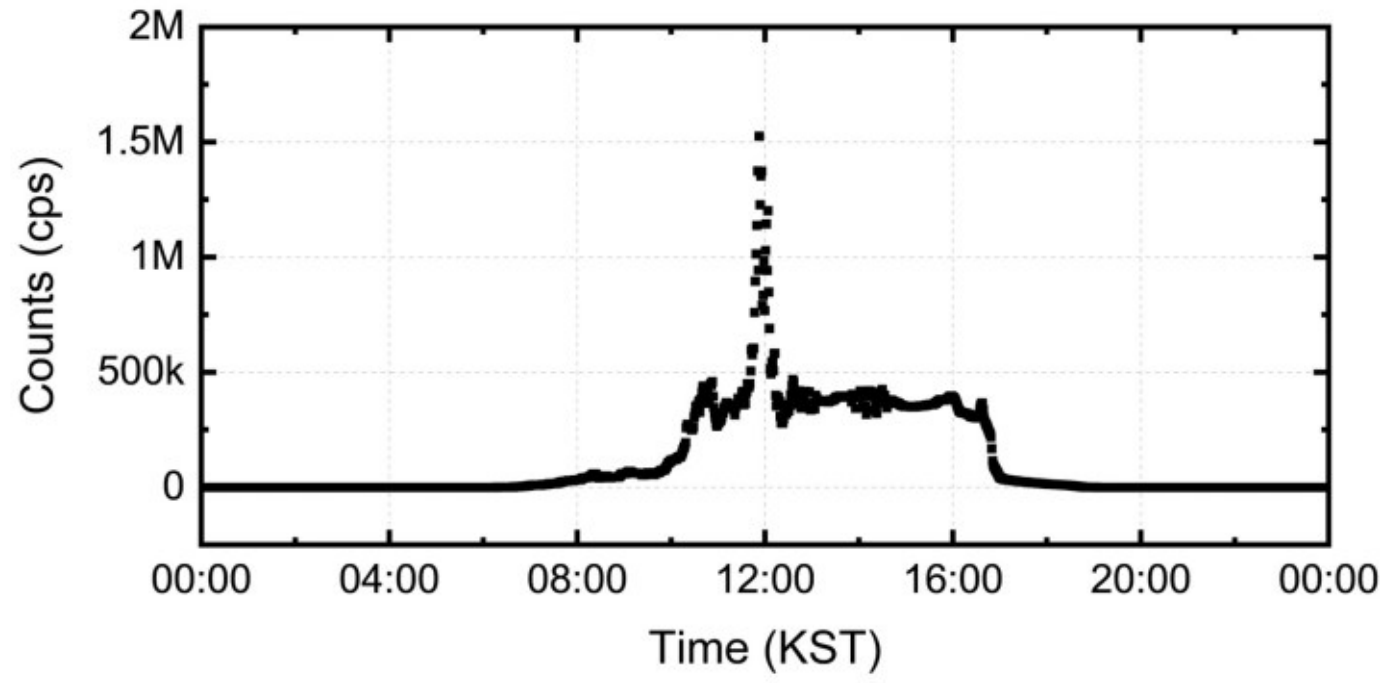

Supplementary Fig. S7. Daylong solar-background singles measured at the receiver optics without the polarization-analysis stage and without receiver filtering. The data provide a reference for the magnitude and temporal variation of the daytime solar background relevant to the outdoor free-space environment.

5.2 Long-term coupling stability with dual-FSM alignment

To support stable free-space-to-fiber coupling in the rooftop experiment, the outdoor link employed active beam stabilization using two fast steering mirrors (FSMs). As an auxiliary verification of link stability, the coupling output was measured with classical light over approximately 2 h under the same dual-FSM alignment configuration. During the measurement period, the relative standard deviation of the received optical power was approximately 2%, indicating that the stabilization scheme maintained a highly stable coupling condition over long durations.

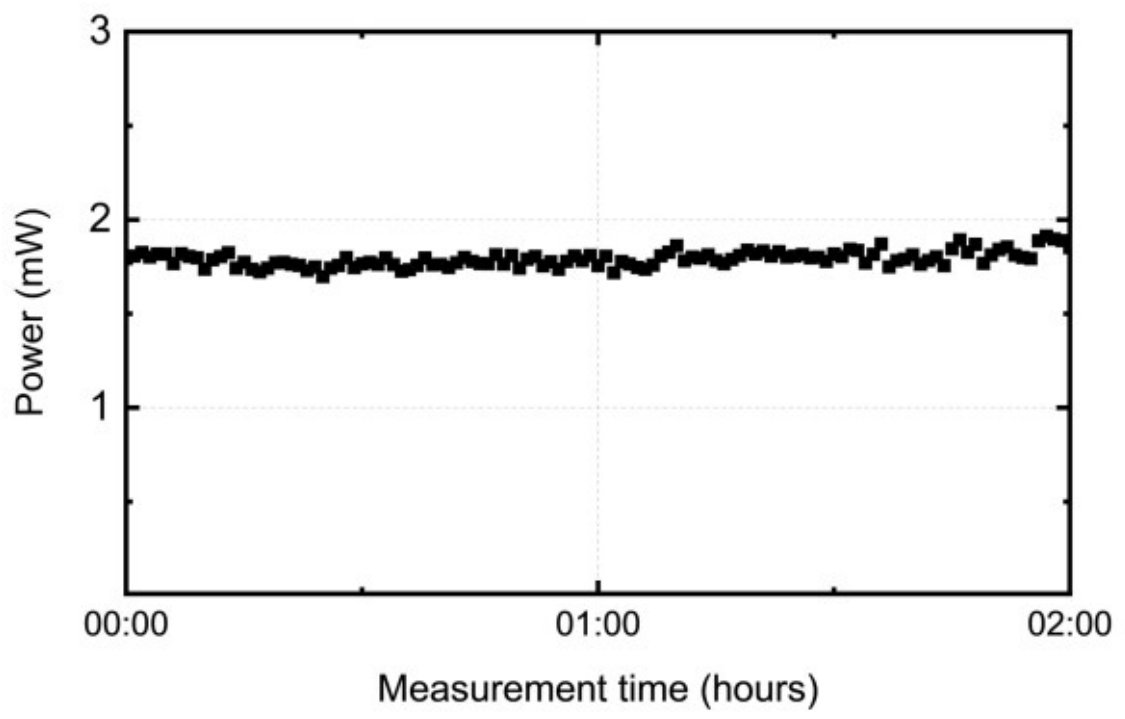

Supplementary Fig. S8. Long-term measurement of the received classical coupling output under the dual-FSM alignment configuration used in the rooftop experiment. The data, acquired over approximately 2 h, demonstrate stable free-space-to-fiber coupling enabled by active beam stabilization with two fast steering mirrors. Over the measurement period, the relative standard deviation of the received optical power was approximately 2%.

## Supplementary References